\documentclass[12pt]{article}
\usepackage{amsmath}
\usepackage{graphicx}
\usepackage{rotating}
\usepackage{natbib}
\usepackage{enumerate}
\usepackage{url}

\newcommand{\blind}{1} 

\usepackage[utf8]{inputenc}
\usepackage[ruled,vlined]{algorithm2e}
\usepackage{bm}
\usepackage{psfrag,epsf}
\usepackage{caption}
\usepackage{subcaption}
\usepackage{float}
\usepackage{ amssymb }
\usepackage[table,xcdraw]{xcolor}
\usepackage{enumitem}
\usepackage{hyperref}
\usepackage{url}
\usepackage{scalerel}
\usepackage{lscape}
\usepackage{booktabs}

\DeclareMathOperator*{\argmax}{arg\,max}

\begin{document}

\def\spacingset#1{\renewcommand{\baselinestretch}%
{#1}\small\normalsize} \spacingset{1}


\if1\blind
{
  \title{\bf Bayesian Structure Learning in Undirected Gaussian Graphical Models: Literature Review with Empirical Comparison}
  \author{Lucas Vogels, Reza Mohammadi, Marit Schoonhoven and Ş. İlker Birbil\\
    Faculty of Economics and Business, University of Amsterdam}
  \maketitle
} \fi

\if0\blind
{
  \bigskip
  \bigskip
  \bigskip
  \begin{center}
    {\LARGE\bf Bayesian Structure Learning in Undirected Gaussian Graphical Models: a Literature Review with Empirical Comparison}
\end{center}
  \medskip
} \fi

\bigskip
\begin{abstract}
\noindent Gaussian graphical models provide a powerful framework to reveal the conditional dependency structure between multivariate variables. The process of uncovering the conditional dependency network is known as structure learning. Bayesian methods can measure the uncertainty of conditional relationships and include prior information. However, frequentist methods are often preferred due to the computational burden of the Bayesian approach. Over the last decade, Bayesian methods have seen substantial improvements, with some now capable of generating accurate estimates of graphs up to a thousand variables in mere minutes. Despite these advancements, a comprehensive review or empirical comparison of all recent methods has not been conducted. This paper delves into a wide spectrum of Bayesian approaches used for structure learning and evaluates their efficacy through a comprehensive simulation study. We also demonstrate how to apply Bayesian structure learning to a real-world data set and provide directions for future research. This study gives an exhaustive overview of this dynamic field for newcomers, practitioners, and experts.
\end{abstract}

\vfill

\newpage

\newpage

\spacingset{1.9} 

\section{Introduction}
\label{sec:intro}

 One can depict conditional dependencies between a large number of variables using undirected graphical models. Using data to recover the structure of a graphical model is called \textit{structure learning}. There are many applications of undirected graphical models. In biology, they are used to recover gene networks \citep{Bhadra2013,GraphHorseshoe,xChandra2021}; in neuroscience, graphical models illustrate the connectivity of the brain \citep{Hinne2014,Alzheimer}; in economics, they map relationships between purchases of customers \citep{economicsapplication}; in finance, they discover how risks of financial institutions are related \citep{SSWang}; and, in psychology, they map relationships between psychological variables \citep{Epskamp2018,psyBSL}.

  Within structure learning, Bayesian methods offer several advantages over frequentist approaches. First, Bayesian methods integrate prior beliefs, which can enhance the precision of learning the structure. For example, one can have an idea about the risk factors of a disease before seeing any patient data. Second, Bayesian methods estimate the entire probability distribution of parameters, whereas frequentist methods only provide point estimates. This allows Bayesian methods to incorporate model uncertainty. Despite these advantages, frequentist methods in structure learning are more popular than Bayesian methods due to their computational efficiency and simplicity. However, in the last decade, new Bayesian methods were proposed that offer computational efficiency and accuracy, even for large-scale graphs. Meanwhile, the development of several software packages makes Bayesian structure learning accessible and practical; see, for example, the R packages \textsf{BDgraph} \citep{BDgraph}, \textsf{ssgraph} \citep{ssgraph}, \textsf{BGGM} \citep{BGGM} and \textsf{baygel} \citep{baygel}.

  Despite the recent developments in Bayesian structure learning, there is no comprehensive review or empirical comparison of all methods. For newcomers to the field, this review provides a clear introduction. For practitioners, we discuss typical challenges when applying Bayesian methods to real data and show the benefits of Bayesian structure learning in practice. For experts, we provide an in-depth overview of Bayesian methods coupled with an extensive simulation study.  

 In this article, we consider general undirected graphs. Moreover, we assume Gaussian data, \textit{i.e.}, we work with Gaussian graphical models (GGMs). GGMs are the most widely researched undirected graphical models due to their simplicity and variety of applications. The focus of this review is on Bayesian methods in GGMs. Besides, we do include a short overview of relevant frequentist methods as they form the starting point for some of their Bayesian counterparts. Moreover, the simulation study contains a frequentist approach to show how Bayesian methods perform relative to the better-known frequentist alternative. We focus on the Bayesian methods of the last two decades since these contributed most to the progress in Bayesian structure learning. GGMs form the foundation of several sub-fields that we briefly touch upon too.
 
Section \ref{sec:problem_intro} provides a general introduction to frequentist and Bayesian structure learning in GGMs. Section \ref{sec:methods} provides a review of the recent literature, and Section \ref{sec:simulation} contains a simulation study comparing the performance of seven state-of-the-art Bayesian algorithms and one frequentist algorithm. In Section \ref{sec:application}, we apply Bayesian methods to a gene data set and discuss typical challenges that arise in real-world applications of Bayesian structure learning. We end with a short overview of related sub-fields and give future perspectives and recommendations in Section \ref{sec:discussion}.

\section{Structure Learning}
\label{sec:problem_intro}
This section introduces structure learning. We start with the problem and notation in Subsection \ref{sec:prob_not}, discuss frequentist approaches in Subsection \ref{sec:frequentistSL}, and end with an introduction to Bayesian structure learning in Subsection \ref{sec:MCMCsampling}.

\subsection{Problem and Notation}
\label{sec:prob_not}
Let $(X_1,\dots,X_p)$ be a vector of random variables, and $X_{-ij}$ denote this vector without the variables $X_i$ and $X_j$ with $i,j=1,\dots,p$ and $i \not = j$. We say that the variables $X_i$ and $X_j$ are conditionally independent, when
\begingroup\abovedisplayskip=5pt \belowdisplayskip=5pt
\[
P(X_i|X_j,X_{-ij}) = P(X_i|X_{-ij}).
\]
\endgroup
In other words, when the values of the variables $X_{-ij}$ are given, knowing the value of $X_j$ does not change the probability distribution of $X_i$. When two variables $X_i$ and $X_j$ are conditionally (in)dependent, they are also referred to as partially (un)correlated \citep{lauritzen}.

One can depict the conditional dependencies between any pair of variables in an undirected graph $G = (V, E)$. Here, every node in the node set $V = \{1,\dots,p\}$ corresponds to a random variable, and the edge set $E$ is defined by $E = \{(i,j) \in V \times V: i < j \text{ and } X_i \text{ and } X_j \text{ are conditionally dependent} \}$. We slightly abuse notation by saying $(i,j) \in G$ when we mean $(i,j) \in E$. We denote the true graph with $G^*=(V, E^*)$. In general, we do not know the true graph. We only observe a sample of $n$ observations of the variables $X_1,\dots, X_p$. Let $x_{li}$ denote the value of the random variable $X_i$ for observation $l$ with $i=1,\dots,p$ and $l=1,\dots,n$. We denote all observations of the $i$th variable as $\mathbf{X}_i = (x_{1i},\dots,x_{ni})^T$. Lastly, $\mathbf{X} = (\mathbf{X}_1,\dots,\mathbf{X}_p)$ denotes the $n \times p$ matrix containing all observations of all variables. We assume that all observations are drawn from a multivariate normal distribution with mean $\mathbf{0}$ and an unknown $p\times p$ covariance matrix $\mathbf{\Sigma}^*$. That is, $(x_{l1},x_{l2},\dots,x_{lp}) \sim \mathcal{N}(\mathbf{0},\mathbf{\Sigma}^*)$ for all $l =1,\dots,n$. 

 The problem of finding the graph $G^*$ was introduced by \citet{Dempster1972} and coined as covariance selection. It is also referred to as structure learning or graphical model determination. The goal of structure learning is to use the sample $\mathbf{X}$ to infer the properties of $G^*$. Let $\mathbf{K}^* = {\mathbf{\Sigma}^*}^{-1}$ with elements $k^*_{ij}$,  $i,j=1,\dots,p$, be the true and unknown precision matrix or concentration matrix. The precision matrix is a helpful tool in unraveling the structure of the true graph, since
\begin{equation}
    \rho_{ij} = -\frac{k^*_{ij}}{\sqrt{k^*_{ii}k^*_{jj}}},
    \label{eq:partialcorr}
\end{equation}
where $\rho_{ij}$ is the partial correlation between variables $X_i$ and $X_j$. For a proof, see  \cite{lauritzen}. Due to \eqref{eq:partialcorr}, we obtain the following relationship between $\mathbf{K}^*$ and $G^*$:
\begin{equation}
\label{eq:relationKandG}
k^*_{ij} = 0 \iff (i,j) \not\in G^*.
\end{equation}

\subsection {Frequentist Structure Learning}
\label{sec:frequentistSL}
This section gives a short overview of frequentist methods and their connections to their Bayesian counterparts. Most frequentist methods in structure learning for GGMs obtain an estimate $\hat{G} = (V,\hat{E})$ of the true graph $G^*$ by either solving multiple regressions or employing the penalized likelihood \citep[Chapter 2.2]{Liang2023}. 
The regression-based approaches start with the observation that we can express each variable $X_i$ in terms of the other variables $X_{-i}$ as follows:
\begin{equation}
    X_i = \sum_{j \not = i} \beta_{ij}X_j + \epsilon_i, \quad  i=1,\dots,p,
    \label{eq:regression}
\end{equation}
where $\epsilon_i$ is uncorrelated with $X_{-i}$ and $\beta_{ij} = \frac{k_{ij}^*}{k_{ii}^*}$ \citep{lauritzen}. Finding the regression coefficients $\beta_{ij}$ in Equation \ref{eq:regression} is thus enough for structure learning, since $\beta_{ij} = 0 \iff  k_{ij}^* = 0 \iff (i,j) \not\in G^*$. In their neighbourhood selection method \citet{nb_selection} use the Lasso regression by \citet{regression_lasso} to obtain estimates for all regression coefficients by solving a sequence of optimization problems given by
\begin{equation}
    \min_{\beta_{ij}, j \not = i} ||\mathbf{X}_i - \sum_{j \not =i}\beta_{ij}\mathbf{X}_j||^2 + \lambda \sum_{j \not = i} |\beta_{ij}|, \quad i=1,\dots,p,
    \label{eq:BayesianLasso}
\end{equation}
where the first term of the objective function is the residual term and the second term (with the regularization parameter $\lambda \geq 0$) is the $\ell_1$-norm regularization term enabling sparse solutions. An edge $(i,j)$ is then included, if $\beta_{ij} \not = 0$. \citet{freq_reg_consistent} show that this approach is consistent, \textit{i.e.}, it recovers the true graph with probability going to $1$ as $n \to \infty$. Similar approaches have been proposed by \citet{Yuan2010}, \citet{Sun2012}, and \cite{TIGER}, each using different methods to find the coefficients $\beta_{ij}$ in \eqref{eq:regression}. The so-called pseudo-likelihood can be derived from Equation \ref{eq:regression} and is used by several Bayesian methods discussed in Sections and  \ref{sec:MCMCjoint} and \ref{sec:gmethods}. Moreover, the Bayesian regression approaches discussed in Section \ref{sec:regression_algorithms} also use Equation \ref{eq:regression}, but use Bayesian methods to estimate the regression coefficients.

The penalized likelihood approach starts with the observation that the log-likelihood can be written as 
\begin{equation*}
    \log P(\mathbf{X} | \mathbf{K}) = C + \frac{n}{2}\log |\mathbf{K}| - \frac{n}{2}tr(\mathbf{S}\mathbf{K}),
\end{equation*}
where $C$ is a constant, $\mathbf{S} = \frac{1}{n}\mathbf{X^T}\mathbf{X}$ is the sample covariance matrix, and $tr(\mathbf{A})$ and $|\mathbf{A}|$ are the trace and determinant of a square matrix $\mathbf{A}$, respectively. Penalized likelihood methods aim to find the maximum likelihood estimator (MLE), \textit{i.e.}, the positive definite precision matrix $\mathbf{K}$ that maximizes the log-likelihood. However, this MLE is generally not sparse. Penalized likelihood methods therefore add a penalty function $p(\mathbf{K})$ as a regularization term and minimize the negative penalized log-likelihood:
\begin{equation}
    \min_{\mathbf{K} \in \mathcal{K}} -\frac{n}{2}\log |\mathbf{K}| + \frac{n}{2} tr(S\mathbf{K}) + \lambda p(\mathbf{K}),
    \label{eq:glasso}
\end{equation}
where $\mathcal{K}$ is the set of all $p \times p$ positive definite matrices. In the case of $\ell_1$-norm regularization, the penalty term is given by $p(\mathbf{K}) = \sum_{i,j, i \not = j}|k_{ij}|$ and the optimization problem \eqref{eq:glasso} becomes a convex program that \citet{YuanLinfreq} solve using the interior point algorithm \citep{detmax}. More efficient is the block coordinate descent approach introduced by \citet{glasso_Banerjee} and improved by \citet{glasso_Friedman}, who coined it the graphical Lasso (glasso) algorithm. To this day, the glasso algorithm remains the most popular in structure learning and is therefore included in the simulation study in Section \ref{sec:simulation}. Performance bounds of the graphical lasso estimate $\mathbf{\hat{K}}$ have been proposed in the Frobenius norm \citep[Theorem 1]{glasso_bound}, in the element-wise $\ell_{\infty}$ norm \citep[Corollary 1]{glasso_bounds}, spectral norm \citep[Equation 43b]{glasso_bounds} and element-wise $\ell_1$ norm \citet[Theorem 14.1.2]{handbookofgraphicalmodels}. Moreover, \citet[Theorem 2]{glasso_bounds} prove that the graph $\hat{G}$ inferred by $\mathbf{\hat{K}}$ converges in probability to the true graph $G^*$. Several other penalty functions $p(\mathbf{K})$ have been proposed, see, for example, \citet{adaptive_glasso}. The tuning parameter $\lambda$ directly impacts the sparsity of the estimate $\mathbf{\hat{K}}$. Several strategies have been proposed for the choice of $\lambda$, including the extended Bayesian information criterion \citep{EBIC}, the rotation information criterion \citep{huge_package}, and $k$-fold cross validation \citep{kfoldcrossvalidation}. The Bayesian methods discussed in Section \ref{sec:Kmethods} have a direct connection to these frequentists penalized likelihood approaches: their Bayesian estimate of the precision matrix is equal to the one obtained by solving the minimization problem given by \eqref{eq:glasso}.

\subsection {Bayesian Structure Learning}
\label{sec:MCMCsampling}
In contrast to frequentist methods that aim to provide an estimate of the true graph, Bayesian methods strive to uncover the probability distribution of the true graph given the sample $\mathbf{X}$. This is called the posterior distribution and is denoted with $P(G | \mathbf{X})$. This is the probability that, given the data $\mathbf{X}$, the true graph is equal to some graph $G$. It is given by Bayes'  rule:
\begin{equation}
    P(G|\mathbf{X}) = \frac{P(\mathbf{X}|G)P(G)}{P(\mathbf{X})} .
    \label{eq:Bayesrule}
\end{equation}
Here, $P(\mathbf{X} | G)$ is the likelihood and denotes the probability of observing the data $\mathbf{X}$ when the true graph is equal to $G$. For each $G$, the prior $P(G)$ describes the given belief that $G$ is the true graph. Lastly, $P(\mathbf{X})$ is the normalizing constant given by $P(\mathbf{X}) = \sum_{G \in \mathcal{G}} P(\mathbf{X}|G)P(G)$, where $\mathcal{G}$ denotes the set of all graphs with $p$ nodes. To evaluate the likelihood $P(\mathbf{X} | G)$ in Equation \ref{eq:Bayesrule}, one could use
\begin{equation}
    P(\mathbf{X}|G) = \int_{\mathbf{K}\in \mathcal{K}}P(\mathbf{X}|G,\mathbf{K})P(\mathbf{K}|G)d\mathbf{K}.
    \label{eq:marginal_likelihood}
\end{equation}
$P(\mathbf{K} | G)$ is the prior distribution on the precision matrix. For each $G$ and $\mathbf{K}$, it describes the belief that when the true graph is equal to $G$, the precision matrix is equal to $\mathbf{K}$. Often the integral in \eqref{eq:marginal_likelihood} is hard to evaluate. Other Bayesian methods therefore focus on the joint posterior distribution of the true graph and precision matrix given by 
\begin{equation}
P(G,\mathbf{K} | \mathbf{X}) = \frac{P(\mathbf{X}|\mathbf{K},G)P(\mathbf{K} | G)P(G)}{P(\mathbf{X})}.
\label{eq:jointposterior}
\end{equation}
Here, $P(\mathbf{X}|\mathbf{K}, G)$ is the joint likelihood and denotes the probability density of $n$ samples from a multivariate normal distribution with mean $\mathbf{0}$ and covariance matrix $\mathbf{K}^{-1}$. It is given by 
\begin{equation}
P(\mathbf{X}|\mathbf{K},G) = (2\pi)^{-pn/2} |\mathbf{K}|^{n/2} \exp \left\{ -\frac{n}{2} tr( \mathbf{K}\mathbf{S} ) \right\}.
\label{eq:likelihood1}
\end{equation}

\section{Review of Methodology}
\label{sec:methods} 
The history of Bayesian structure learning is one of progress. From inefficient approaches restricted to small and decomposable graphs in the 1990s and 2000s to the efficient, accurate, and generally applicable methods of today. This chapter presents the story of this progress. We briefly cover the most relevant approaches before 2010 before we give a detailed outline of all recent methods. 

Bayesian approaches in structure learning aim to uncover the posterior distribution $P(G|\mathbf{X})$ of the graph. For that, one could either evaluate the posterior \eqref{eq:Bayesrule} or the joint posterior \eqref{eq:jointposterior}. For both, the prior distribution $P(\mathbf{K}|G)$ is required. Early Bayesian structure learning focuses on proposing this prior. Ideally, this prior is conjugate, \textit{i.e.}, the prior and posterior follow the same distribution family. \citet{Gwishart1993} define such prior and call it the Hyper Inverse-Wishart (HIW) prior. Using the HIW prior, \citet{decomposable2} and \citet{decomposable1} design algorithms for Bayesian structure learning that output the posterior for small graphs.  The HIW prior, however, is only defined on decomposable graphs, which form a small subset of the whole graph space. \citet{Gwishart2002} solves this by extending the HIW to general graphs, a distribution that \citet{Massam2005} later coin the $G$-Wishart distribution. Its density is given by 
\begin{equation}
P(\mathbf{K}| G)=\frac{1}{I_G(b,\mathbf{D})} |\mathbf{K}|^{\frac{b}{2}-1} \exp \left\{-\frac{1}{2} tr( \mathbf{K} \mathbf{D} ) \right\}I(\mathbf{K} \in \mathcal{P}_G),
\label{eq:Gwishart}
\end{equation}
where $\mathcal{P}_G \subset \mathcal{K}$ denotes the set of symmetric positive definite matrices $\mathbf{K}$ that have $k_{ij} = 0$ when $(i,j) \not \in G$. Here, $I(\mathbf{K} \in \mathcal{P}_G)$ is an indicator function that is equal to one, if $\mathbf{K} \in \mathcal{P}_G$, and zero otherwise. The symmetric positive definite matrix $\mathbf{D}$ and the scalar $b>2$ are called the scale and shape parameters of the $G$-Wishart distribution. They are mostly set to $b=3$ and $\mathbf{D}=\mathbf{I}$, where $\mathbf{I}$ is the p-dimensional identity matrix. The normalizing constant
\begin{equation}
I_G(b,\mathbf{D}) = \int_{\mathbf{K} \in \mathcal{P}_G} |\mathbf{K}|^{\frac{b-2}{2}} \exp \{-\frac{1}{2} tr( \mathbf{K} \mathbf{D} ) \}d\mathbf{K}
\label{eq:normconst}
\end{equation}
ensures that $\int_{\mathbf{K} \in \mathcal{P}_G} P(\mathbf{K}| G) d\mathbf{K} = 1$. \citet{recent_Gwish} present techniques to evaluate the normalising constant for certain types of graphs. For general graphs, however, the normalizing constant is troublesome to evaluate. \citet{Gwishart2002}, \cite{Massam2005}, and \cite{Stochastic2013} propose approximations. Using the $G$-Wishart prior, the marginal likelihood \eqref{eq:marginal_likelihood} becomes
\begin{equation}
P(\mathbf{X}|G)  = (2\pi)^{-np/2} \frac{I_G(b+n,\mathbf{D}+\mathbf{X}^T\mathbf{X})}{I_G(b,\mathbf{D})}.
\label{eq:ylikelihood2}
\end{equation}
Using the approximations for the normalizing constants in this ratio, one can approximate $P(\mathbf{X}|G)$. \citet{Stochasticsearch2005} and \citet{Stochastic2013} both design stochastic search algorithms that in each iteration favour a move to a graph with a high posterior probability $P(G|\mathbf{X}) \propto P(\mathbf{X}|G)P(G)$. However, the approximation of the normalizing constants in \eqref{eq:ylikelihood2} is unstable and computationally expensive, limiting structure learning to small graphs with less than 20 nodes.

That is why others focus on inference on the joint posterior \eqref{eq:jointposterior}. For this and for inference on the precision matrix in general, it is required to sample from the $G$-Wishart distribution \eqref{eq:Gwishart}, which is challenging due to the normalizing constant \eqref{eq:normconst}. This led to a string of articles on efficient sampling from the $G$-Wishart distribution, see \citet{Wang2012} for a review of those. However, sampling from the $G$-Wishart distribution remained computationally expensive.

The period after 2010 saw several discoveries accelerating Bayesian structure learning. They include solutions to the challenges discussed before: computationally efficient approximation of the normalizing constant \eqref{eq:normconst} and computationally efficient $G$-Wishart sampling. They also contain entirely new perspectives and approaches, enabling the creation of new methods that push the boundaries of accuracy and computational efficiency. These approaches form the topic of this chapter. We group them into four different categories: Markov chain Monte Carlo (MCMC), Expectation Conditional Maximization (ECM), regression, and hypothesis approaches. The most common Bayesian methods in each category are listed in Table \ref{tab:searchliterature}. Next, we provide an overview of the workings and theoretical properties of these methods.

\renewcommand{\arraystretch}{0.6}
\begin{table}[h!]
\begin{tabular}{ l  l l c}
\arrayrulecolor{blue}\toprule
Category & Method name & Reference & \small Included in sim.\\
\hline

MCMC on $\mathcal{G} \times \mathcal{K}$ & RJ - O & \cite{RJ} & \\
 &    RJ - WL & \cite{Wang2012} &\\
 &   RJ - CL & \cite{Cheng2012} &\\
 &   RJ - D & \cite{DRJ} &\\
 &     RJ - DCBF & \cite{Hinne2014} &\\
 &     RJ - A & \cite{BDgraph} &\\
 &    RJ - WWA & \cite{WWA} & \checkmark\\
  &    RJ - UB & \cite{SMCBoom} & \\
 &   BD - O & \cite{BDMCMC1} & \\
 &     BD - A & \cite{Accelerating} & \checkmark\\
&     SS - O & \cite{SSWang} & \checkmark \\     
& SS - AT & \cite{ATCHADE} &  \\
& SS - BC & \cite{CONCORD}* &  \\
MCMC on $\mathcal{G}$  & G - Stingo & \cite{Stingo} &\\ 
& G - MPLRJ & \cite{MPLpaper}* &\\
  & G - MPLBD & \cite{MPLpaper}* & \checkmark\\

MCMC on $\mathcal{K}$ & K - BGLasso1 & \cite{BayGraphicalLasso} & \\
 &  K - BGLasso2 & \cite{BCLasso} & \\
 & K - RIW & \cite{RIW} &  \\
 & K - Horseshoe & \cite{GraphHorseshoe} &  \checkmark \\
& K - LRD & \cite{xChandra2021}*  & \\
& K - BAGR & \cite{Smith2023}*  &\\
& K - Horseshoe-like & \cite{HorseshoeLike} &  \\
\hline
ECM & ECM - EMGS & \cite{EMGS} &  \\
 &   ECM - BAGUS & \cite{SSLasso} &  \checkmark\\
 &   ECM - HL & \cite{HorseshoeLike} &  \\
\hline
Regression & R - BLNRE & \cite{BLNRE} & \\
& R - P & \cite{Projection} & \\
\hline
Hypothesis & H - LR & \cite{Leday} & \\
& H - BGGM & \cite{Williams2020} & \checkmark \\
 \bottomrule
\end{tabular}
\caption{\textit{Overview of all the state-of-the-art methods for Bayesian structure learning in Gaussian Graphical Models. They are listed per category and chronologically. The last column indicates whether the method is included in the simulation study in Section \ref{sec:simulation}. Papers marked with * have not yet been published.}}
\label{tab:searchliterature}
\end{table}

\subsection{MCMC-based Approaches}
\label{sec:MCMC}
This section gives a short introduction to Markov chain Monte Carlo (MCMC) sampling before describing the main MCMC-based approaches in Bayesian structure learning. 

Most methods in Bayesian structure learning use MCMC sampling, because of the variety of inferences it provides. Specifically, when using MCMC sampling, one can approximate the expected value of $h(G^*)$ for any information function $h: \mathcal{G} \to \mathbb{R}$. For example, one can choose $h(G)$ equal to $1$ when an edge $e$ is in $G$, and $0$ otherwise. This will give the posterior edge inclusion probabilities
\begin{align}
    p_{e} & := P(e \in G^* | \mathbf{X}) = E((h(G^*)|\mathbf{X}). \label{eq:edge_inclusion_prob} 
\end{align}
Similarly, with other information functions $h$, one can evaluate different posterior probabilities. For example, the probability that the true graph will have more than $10$ edges, the probability that the true graph is connected, and so on. However, this flexibility comes at a cost: MCMC algorithms are computationally demanding because they iteratively explore large parameter spaces. 

Let $P(\mathbf{X}|\bm{\theta})$ be the likelihood conditional on some parameter vector $\bm{\theta} \in \mathcal{T}$. Let $\bm{\theta}^*$ be the true and unknown parameter vector. We want to perform inference on the posterior $P(\bm{\theta}|\mathbf{X})$. We are interested in obtaining $E(h(\bm{\theta}^*)|\mathbf{X}) = \int_{\bm{\theta}} h(\bm{\theta})P(\bm{\theta}|\mathbf{X})d\bm{\theta}$
for some information function $h:\mathcal{T} \to \mathbb{R}$. MCMC algorithms start at an initial realization of the parameter vector $\bm\theta^{(0)}$ and then move to $\bm\theta^{(1)}$, $\bm\theta^{(2)}$, all the way to $\bm\theta^{(S)}$ with $S \in \mathbb{N}$. We call the resulting sequence $(\bm\theta^{(0)},\dots,\bm\theta^{(S)})$ a Markov chain. For every $s = 1,\dots,S$, $\bm\theta^{(s)}$ is called a state of the Markov chain. All states are elements of the state space $\mathcal{T}$. The transition kernel $P(x, A)$ denotes the probability that the next state will be part of the set $A \subset \mathcal{T}$, given that the current state is $x \in \mathcal{T}$. A suitable transition kernel ensures that the chain converges to the so-called stationary distribution. There are three conditions sufficient for this convergence \citep[Theorem 1]{stationary_distr_exists}, the most crucial being the balance condition, which is often replaced by a stricter condition called the detailed balance condition, see Equation 1 in \citet{Green1995}. Now, if the balance condition holds and if the stationary distribution is equal to the posterior distribution, the distribution of our Markov Chain will get arbitrarily close to the posterior as we increase the chain length $S$. This holds no matter the starting position  $\bm\theta^{(0)}$ of our Markov Chain. We can then get arbitrarily close to our desired expected value using
\begin{equation}
    E(h(\bm\theta^*)| \mathbf{X}) = \lim_{S \to \infty} \frac{1}{S}\sum_{s=1}^S h(\bm\theta^{(s)}) \approx  \frac{1}{S}\sum_{s=1}^S h(\bm\theta^{(s)}).
    \label{eq:model_average_MCMC}
\end{equation}

\subsubsection{MCMC-based approaches on the joint space}
\label{sec:MCMCjoint}
Most of the MCMC-based approaches in Bayesian structure learning create a Markov Chain $((G^{(0)},\mathbf{K}^{(0)}),\dots,(G^{(S)},\mathbf{K}^{(S)}))$ over the joint space of graphs and precision matrices that converges to the joint posterior \eqref{eq:jointposterior}. Using Equation \ref{eq:model_average_MCMC} these methods can perform inference on almost all characteristics of both the graph and the precision matrix. We call these methods joint MCMC methods. They come in three types: reversible jump (RJ), birth-death (BD), and spike-and-slab (SS) methods. All joint MCMC methods are listed in Table \ref{tab:jointMCMCliterature} at the end of this subsection.

RJ methods are the first MCMC-based methods in Bayesian structure learning. They use the $G$-Wishart prior given by \eqref{eq:Gwishart}. RJ methods are a variant of Metropolis-Hastings methods that were introduced by \citet{Green1995}. Algorithm \ref{alg:RJ} represents the pseudo-code for this class of methods. 

\begin{algorithm}[H]
\caption{Reversible Jump (RJ) algorithm}
\label{alg:RJ} 
 \KwIn{ Data $\mathbf{X}$ and an initial graph $G^{(0)}$. }
 \For{state $s=1,\dots,S$}{ 
   Set $G = G^{(s-1)}$\;
   Propose a new graph $G'$ by adding or deleting one edge from $G$\;
   Accept the proposal with acceptance probability $\alpha(G',G)$\;
   Set $G^{(s)}=G'$ if accepted, set $G^{(s)}=G$ if rejected\;
   Sample $\mathbf{K}^{(s)}$ from the $G^{(s)}$-Wishart distribution\;
 }
 \KwOut{Samples $(G^{(0)},\mathbf{K}^{(0)}),\dots,(G^{(S)},\mathbf{K}^{(S)})$ from the joint posterior \eqref{eq:jointposterior}} 
\end{algorithm}
One can define the acceptance probabilities $\alpha(G', G)$ such that the balance condition holds and the resulting MCMC chain converges to the joint posterior. The first three RJ methods to use this strategy are the RJ-Original (RJ-O) method by \citet{RJ}, the RJ-Wang Li (RJ-WL) method by \citet{Wang2012}, and the RJ-Cheng Lenkoski (RJ-CL) method by \citet{Cheng2012}. These methods form the foundation for all later reversible jump methods. Their approach to sample from the $G$-Wishart distribution, however, is not scalable to large-scale graphs and lacks a theoretical guarantee of convergence. To address these issues, \citet{DRJ} introduces an efficient exact sampler from the $G$-Wishart distribution. The resulting RJ-Double (RJ-D) method thereby becomes the first joint method to achieve detailed balance, and hence, theoretical convergence. \citet{Hinne2014} combine several existing techniques to improve the overall efficiency in the RJ-Double Continuous Bayes Factor (RJ-DCBF) method. Despite these improvements, RJ methods are still computationally demanding for graphs with more than 50 variables. One of the reasons is that the acceptance probabilities contain a ratio of normalizing constants that is time-consuming to evaluate. \citet{Accelerating} tackles this by developing a closed-form approximation of the ratio used in the RJ - Approximation (RJ-A) method by \citet{BDgraph_paper}. Due to the approximation, the RJ-A method loses theoretical convergence but gains computational efficiency. The RJ-Weighted Proposal (RJ-WWA) method by \citet{WWA} combines the best of both worlds: theoretical convergence and computational efficiency. It achieves this with two main improvements. First, the RJ-WWA algorithm is more likely to propose graphs with higher acceptance probabilities thereby improving the low acceptance probabilities of earlier algorithms. Second, they introduce a pre-acceptance step, in which the acceptance probabilities are approximated using the computationally efficient approximation of \citet{Accelerating}. Only when pre-accepted, the exact acceptance probabilities are calculated. The most recent reversible jump method is the RJ-Unbiased (RJ-UB) method by \citet{SMCBoom}. The RJ-UB method introduces unbiased estimation to Bayesian structure learning by applying a technique of \citet{glynn_rhee}. It requires the construction of two Markov Chains $(G^{(0)},\dots,G^{(S)})$ and $(\bar{G}^{(0)},\dots,\bar{G}^{(S)})$ that meet at some time $\tau < \infty$, that is, $G^{(s)} = \bar{G}^{(s)}$ for all $s \geq \tau$. Now, for any $q=1,2,\dots$, the quantity 
\begin{equation}
    \hat{h}_q  = h(G^{(q)}) + \sum_{s=q+1}^\tau [h(G^{(s)}) - h(\bar{G}^{(s)})]
    \label{eq:unbiased_estimator}
\end{equation}
is an unbiased estimator of $E(h(G^*)|\mathbf{X})$. To create the two Markov chains, \citet{SMCBoom} use a combination of a sequential Monte Carlo (SMC) method and a particle-independent Metropolis-Hastings method.  \citet{SMCBoom} are not the first to use SMC techniques in Bayesian structure learning. \citet{SMCTan} also use an SMC sampler. Their method, however, is computationally cumbersome and comes with no theoretical guarantees. 

\citet{BDMCMC1} develop a different strategy to tackle the issue of low acceptance probabilities and create a new class of joint MCMC methods: Birth-Death (BD) methods. Their BD-Original (BD-O) method abolishes acceptance probabilities altogether while maintaining theoretical convergence. The BD-O method designs a continuous time Markov chain \citep{Cont_time_mcmc}. This chain spends a continuous time $W(s)$ in each state $(G^{(s)},K^{(s)})$ before moving to the next state. The next state is determined by adding or removing an edge based on so-called birth-death rates that are recalculated at every iteration. Its pseudo-code is shown in Algorithm \ref{alg:BD}.

\begin{algorithm}[H]
\label{alg:BD} 
\caption{Birth-death (BD) algorithm }
 \KwIn{ Data $\mathbf{X}$ and an initial graph $G^{(0)}$.  }
 \For{state $s=1,\dots,S$}{ 
   Calculate the birth-death rate $r_e$ for every edge $e = (i,j)$\;
   Calculate the waiting time $W(s-1) = \frac{1}{\sum_er_e}$\;
   Select one edge $e$ with a probability proportional to its birth-death rate\;
   Obtain $G^{(s)}$ by flipping edge $e$ in $G^{(s-1)}$\;
   Sample $\mathbf{K}^{(s)}$ from the $G^{(s)}$-Wishart distribution\;
 }
 \KwOut{Samples $(G^{(0)},\mathbf{K}^{(0)}),\dots,(G^{(S)},\mathbf{K}^{(S)})$ and waiting times $(W^{(0)},\dots,W^{(S)})$}
\end{algorithm}
The BD-O method meets the balance condition and therefore converges to the joint posterior. The desired expected value can now be obtained using
\begin{equation}
    E(h(G^*,\mathbf{K}^*)| \mathbf{X}) \approx  \frac{1}{W}\sum_{s=1}^S W(s) h(G^{(s)},\mathbf{K}^{(s)}),
\end{equation}
where $W$ is the sum of the waiting times of all states. Per iteration, the BD-O algorithm is computationally costly because it calculates the birth and death rates for all edges. However, its MCMC chain needs less iterations to converge because it moves to a new graph every iteration. Moreover, the birth and death rates can be calculated in parallel. Just like the acceptance probabilities of the RJ methods, the birth and death rates of the BD-O method contain a ratio of normalizing constants. Using the approximation of this ratio by \citet{Accelerating}, the BD-Approximation (BD-A) method achieves a significant reduction in computational cost, but the balance condition no longer holds, and there is no theoretical convergence. 

Despite all the mentioned improvements, RJ and BD methods still suffer from a high computational burden. This has two reasons: first, at every iteration, a new precision matrix needs to be sampled from the $G$-Wishart distribution, which remains, despite the direct sampler of \citet{DRJ}, computationally costly. Second, at every MCMC iteration, the graph changes by at most one edge. This leads to poor mixing and convergence of the MCMC chain, especially for large-scale graphs. Both problems are overcome by a new class of joint MCMC methods called Spike-and-Slab (SS) methods. The key element of SS methods is their prior on the precision matrix, called the SS prior \citep{vanucci}. An SS prior samples each element $k_{ij}$ of the precision matrix individually, such that $k_{ij}$ is likely to be close to zero for $(i,j) \not \in G$ (the spike) and likely to be away from zero for $(i,j) \in G$ (the slab). The elements of the precision matrix are still dependent, since the resulting matrix $\mathbf{K}$ needs to be positive definite. Moreover, like the $G$-Wishart prior, SS priors generally contain an intractable normalizing constant. SS priors allow for the design of efficient algorithms that sample an entirely new graph at each iteration. The Markov chain's mixing is therefore greatly improved compared to the RJ and BD methods. Algorithm \ref{alg:SS} presents the pseudo-code for the class of SS algorithms. 
\begin{algorithm}[H]
\label{alg:SS} 
\caption{Spike-and-slab (SS) algorithm }
 \KwIn{ Data $\mathbf{X}$ and an initial graph $G^{(0)}$.}
 \For{state $s=1,\dots,S$}{ 
   Sample a new precision matrix $\mathbf{K}^{(s)}$\;
   For every edge $e=(i,j)$, calculate the probability $l_e$\;
   Obtain $G^{(s)}$ by including every edge $e$ with probability $l_e$\;
 }
 \KwOut{Samples $(G^{(0)},\mathbf{K}^{(0)}),\dots,(G^{(S)},\mathbf{K}^{(S)})$ from the joint posterior \eqref{eq:jointposterior}} 
\end{algorithm}
In the SS-Original (SS-O) method, \cite{SSWang} 
introduces SS priors to Bayesian structure learning. This prior samples off-diagonal elements $k_{ij}$ from a normal distribution with variance $v_0$ when $(i,j) \not \in G$ and with variance $v_1>v_0$ when $(i,j) \in G$. Diagonal elements are sampled from an exponential distribution with parameter $\eta$. The SS-Atchadé (SS-AT) method by \citet{ATCHADE} and the SS-BCONCORD (SS-BC) by \citet{CONCORD} set off-diagonal elements $k_{ij}$ to zero when $(i,j) \not \in G$. These so-called \textit{discrete} SS priors were introduced by \citet{BLSM}. The SS-AT and SS-BC methods do not use the exact likelihood \eqref{eq:likelihood1}, but instead consider an approximation. These approximations are variants of the pseudo-likelihood approximation \citep{Besag1975} given by
\begin{equation}
    \prod_{i=1}^p P(\mathbf{X}_i|\mathbf{X}_{-i},G,\mathbf{K}) \approx P(\mathbf{X}|\mathbf{K},G), \label{eq:pseudo_likelihood} 
\end{equation}
where $\mathbf{X}_{-i}$ denotes the data matrix $\mathbf{X}$ with the column $\mathbf{X}_i$ removed. Both the SS-AT and SS-BC methods motivate the use of this approximation by proving a theoretical property called posterior contraction. This means that the posterior probability of precision matrices $\mathbf{K}$ that are ``far away" from the true precision matrix $\mathbf{K}^*$ goes to zero, as the number of observations goes to infinity. Or, put differently, the posterior distribution $P(\mathbf{K}|\mathbf{X})$ is concentrated around the true value $\mathbf{K}^*$. The SS-AT and SS-BC methods combine the discrete SS prior with the approximated likelihood to construct computationally efficient MCMC algorithms. On top of that, \citet{CONCORD} prove that their SS-BC method provides a consistent estimate of the graph, \textit{i.e.} as the number of observations goes to infinity, the estimated edge inclusion probabilities $p_e$ \eqref{eq:edge_inclusion_prob} converge (in probability) to $1$ for edges in the true graph and to $0$ for edges not in the true graph. 

\renewcommand{\arraystretch}{0.6}
\begin{table}[ht]
\begin{tabular}{ l  l l l l  }
\arrayrulecolor{blue}\toprule
Prior on $\mathbf{K}$ & Discr/Cont. & Method name & Reference & Conv.\\
\hline
$G$-Wishart & Discrete  & RJ - O & \cite{RJ} & \\
 &  &   RJ - WL & \cite{Wang2012} & \\
 &   & RJ - CL & \cite{Cheng2012} &\\
 &    & RJ - D & \cite{DRJ} &\checkmark \\
 &    & RJ - DCBF & \cite{Hinne2014} &\checkmark \\
 &    & RJ - A & \cite{BDgraph} &\\
 &    & RJ - WWA & \cite{WWA} &\checkmark \\
 &    & RJ - UB & \cite{SMCBoom} &\checkmark \\
 &  Continuous & BD - O & \cite{BDMCMC1} &\checkmark \\
 &     & BD - A & \cite{Accelerating} &\\
Spike-and-slab & Discrete  & SS - O & \cite{SSWang} &  \\
& & SS - AT & \cite{ATCHADE} &  \\
& & SS - BC & \cite{CONCORD}* &  \\

 \bottomrule
\end{tabular}
\caption{\textit{Overview of joint MCMC methods for Bayesian structure learning in GGMs with their corresponding prior on $\mathbf{K}$, type of Markov chain (discrete or continuous), name, and reference. The last column indicates whether an algorithm theoretically converges to the posterior distribution. Papers marked with * have not yet been published.}}
\label{tab:jointMCMCliterature}
\end{table}
\subsubsection{MCMC-based approaches on the graph space}
\label{sec:gmethods}
A benefit of joint MCMC methods is that with the samples $(G^{(0)},\mathbf{K}^{(0)}),\dots,(G^{(S)},\mathbf{K}^{(S)})$, one can perform inference on the precision matrix. However, one is often just interested in retrieving the structure of the true graph. In that case, obtaining a new precision matrix at every iteration is a computational burden. The class of methods in this subsection overcomes this burden by creating a Markov chain $(G^{(0)},\dots,G^{(S)})$ only over the graph space. We also refer to these types of methods as G-methods. 

\citet{Stingo} introduce this strategy in their G-Stingo method. If a graph $G_d$ is decomposable, the pseudo-likelihood approximation \eqref{eq:pseudo_likelihood} becomes exact. \citet{Stingo} use this fact to design an MCMC algorithm over the space of decomposable graphs. They then design a framework to generalize this chain to the space of all graphs $\mathcal{G}$.

The pseudo-likelihood approximation \eqref{eq:pseudo_likelihood} can also be used to approximate the marginal likelihood $P(\mathbf{X}|G)$: 
\begin{equation} 
\begin{split}
P(\mathbf{X}|G) & = \int_{\mathbf{K}}P(\mathbf{X}|G,\mathbf{K})P(\mathbf{K}|G)d\mathbf{K} \\
& \approx \int_{\mathbf{K}}\prod_{i=1}^p P(\mathbf{X}_i|\mathbf{X}_{-i},\mathbf{K},G)P(\mathbf{K}|G)d\mathbf{K} \\
& := \hat{P}(\mathbf{X}|G), \\
\end{split}
\label{eq:MPL}
\end{equation}
where $\hat{P}(\mathbf{X}|G)$ is called the marginal pseudo-likelihood (MPL). \citet{PseudoPensar} introduce the concept of marginal pseudo-likelihood to GGMs. They use the Wishart prior with scale parameter $b=p$ and $D = \mathbf{S}$. The Wishart prior is equivalent to the $G$-Wishart prior \eqref{eq:Gwishart}, where $G$ is the complete graph. Note that this prior uses the data $\mathbf{X}$. Such priors are called fractional priors. \citet{PseudoPensar} choose this prior because it leads to a closed-form expression of $\hat{P}(\mathbf{X}|G)$. They design a hill-climbing algorithm that, at each iteration, moves to a graph with a higher marginal pseudo-likelihood.

The G-MPL-Reversible Jump (G-MPLRJ) method by \citet{MPLpaper} combines the closed-form expression of $\hat{P}(\mathbf{X}|G)$ with an MCMC algorithm to sample from the so-called pseudo-posterior $\hat{P}(G | \mathbf{X}) := \hat{P}(\mathbf{X}|G)P(G)/P(\mathbf{X})$. Their discrete time MCMC algorithm creates the chain $(G^{(0)},\dots,G^{(S)})$.  The pseudo-code resembles Algorithm \ref{alg:RJ} with two differences. First, no precision matrix is sampled at every iteration. Second, the acceptance probabilities $\alpha(G',G)$ are calculated using the marginal pseudo-likelihood, greatly improving the computational efficiency. The same work also presents the G-MPL-Birth Death (G-MPLBD) method. Again using the approximation in \eqref{eq:MPL}, this method designs a continuous time MCMC similar to Algorithm \ref{alg:BD}. Both the G-MPLRJ and the G-MPLBD algorithm meet the balance conditions and therefore converge to the pseudo-posterior. Similar to the SS - BC by \citet{CONCORD} (see Subsection \ref{sec:MCMCjoint}), \citet{MPLpaper} prove two theoretical results: the pseudo-posterior concentrates around the true graph $G^*$ and the edge inclusion probabilities are consistent.

\subsubsection{MCMC-based approaches on the precision matrix space}
\label{sec:Kmethods}
Methods on the space of precision matrices remove the necessity of sampling a graph at every iteration. They create an MCMC chain $(\mathbf{K}^{(1)},\dots,\mathbf{K}^{(S)})$ solely over the space of positive definite $p \times p$ matrices. The methods are computationally efficient and have appealing theoretical properties. We also refer to these types of methods as K-methods. K-methods do not use a prior on the graph and are, therefore, not able to incorporate a prior belief in the graphical structure. Moreover, since K-methods do not provide MCMC samples in the graph space, they still require some operation to select a graph $G$ from the posterior samples of the precision matrix. Two examples of such operations are thresholding and credible intervals. The thresholding method \citep{BayGraphicalLasso} first computes the edge inclusion probability $p_{e}$ of an edge $e$ and then includes it in the estimate of the graph if $p_e > \gamma$ where $\gamma \in (0,1) $ is some threshold.  The credible intervals method \citep{GraphHorseshoe} first constructs intervals, for every edge $(i,j)$, that contains $\gamma \%$ of the values $\{k_{ij}^{(1)},k_{ij}^{(2)},\dots,k_{ij}^{(S)}\}$ for some user-defined $\gamma$. An edge is then excluded if and only if the interval includes zero. 

For K-methods, the prior on the precision matrix $P(\mathbf{K} | G)$ no longer depends on $G$ and can therefore be written as $P(\mathbf{K})$. We call these priors K-only priors. Most K-only priors are defined by two probability distributions. One for off-diagonal elements and one for diagonal elements. The elements of K-only priors are not independent due to the positive definite constraint and these priors also contain an intractable normalizing constant. Specific choices of K-only priors enable algorithms to be computationally efficient and/or have attractive theoretical properties. 

All K-methods share one or more of three useful theoretical properties. First, all K-methods meet the balance condition and therefore converge to the posterior distribution of the precision matrix \citep[Theorem 1]{stationary_distr_exists}. Second, most K-methods have a connection with the frequentist graphical lasso algorithm. Specifically, priors of K-methods are often chosen such that the precision matrix that maximizes the posterior $P(\mathbf{K}|\mathbf{X})$ is equal to the graphical lasso estimate obtained by solving the optimization problem \eqref{eq:glasso} for some regularization parameter $\lambda$ and penalty function $p(\mathbf{K})$. The third and last theoretical property is the concentration of the precision matrix around the true value as in the SS-BC algorithm by \citet{CONCORD} (see also Section \ref{sec:MCMCjoint}). Table \ref{tab:kmethods} contains all K-methods with their corresponding K-only priors and theoretical properties.

\begingroup
\setlength{\tabcolsep}{4pt} 
\begin{table}[h]
\begin{tabular}{l l l l l l l l}
\hline
Reference &  Name & \multicolumn{2}{c}{$P(k_{ij}) $} & & \multicolumn{3}{c}{Theor. Prop.}
\\
\cline{3-4} \cline{6-8}
& &   $i \not = j$     & $i = j$ & &i & ii & iii    \\
\hline

\cite{BayGraphicalLasso} & K - BGLasso1 & Double exp. &  Exp.  & &\checkmark & \checkmark &  \\

\cite{BCLasso} & K - BGLasso2 & Double exp. &  Exp. & &\checkmark & \checkmark & \\

\cite{RIW} & K - RIW & Wishart & Wishart & &\checkmark  & & \\ 

\cite{GraphHorseshoe} & K - Horseshoe & Horseshoe & $\propto 1$ & &\checkmark  & & \\

\cite{xChandra2021}* & K - LRD & NA & NA & & \checkmark &  & \checkmark \\

\cite{Smith2023}* & K - BAGR & Norm.  &  Trunc. norm.  & & \checkmark & \checkmark &  \\

\cite{HorseshoeLike} & K - Horseshoe-like & Horseshoe-like & $\propto 1$ & & \checkmark & \checkmark & \checkmark \\
\hline
\end{tabular}
\caption{\textit{K-methods for Bayesian structure learning with their corresponding reference, name, K-only prior, and theoretical properties. Note that all marginal K-only priors are not equal to the distributions mentioned in the table due to the positive definite constraint. The horseshoe and horseshoe-like distribution are defined in the corresponding articles. The theoretical properties are i) the Markov chain converges to the posterior distribution, ii) the maximum a posteriori estimate equals the maximum likelihood estimate \eqref{eq:glasso}, and iii) the posterior concentrates around the true value. Papers with a $^*$ are not yet published.}}
\label{tab:kmethods}
\end{table}
\endgroup

K-methods are not scale-invariant, \textit{i.e.} scaling the observations of one or more variables affects the posterior distribution.  K-methods typically deal with this by standardizing the data to have unit variance. \citet{partial_corr_glasso} observe, however, that this standardization can adversely affect inference. Instead, \citet{partial_corr_glasso} propose to perform inference directly on the partial correlations and define a class of priors for which the resulting posterior is scale-invariant. 

\subsection{Expectation Conditional Maximization Approaches}
\label{sec:EM_methods}
MCMC-based approaches are popular due to the wide variety of inferences they can provide using Equation \ref{eq:model_average_MCMC}. They are, however, computationally demanding because of the large parameter space they explore. Expectation Conditional Maximization (ECM) methods are a more efficient alternative. They allow for the quick computation of the maximum a posteriori (MAP) estimate of the precision matrix but do not provide inference on the rest of the posterior. The three ECM methods in Bayesian structure learning are listed in Table \ref{tab:ECMmethods}.

At every iteration $s=1,\dots,S$, an ECM algorithm renders a new precision matrix $\mathbf{K}^{(s)}$. The sequence $(\mathbf{K}^{(1)},\mathbf{K}^{(2)},\dots)$ converges to the MAP estimator. In each iteration two steps are applied: the Expectation step (E-step) and the Conditional-Maximization step (CM-step).  In iteration $s$, the E-step evaluates $Q(\mathbf{K}|\mathbf{K}^{(s-1)})$, defined as the expected value of the log posterior with respect to the current conditional distribution of $G$ given $\mathbf{K}$ and $\mathbf{X}$. In the CM-step the new precision matrix $\mathbf{K}^{(s)}$ is set to the matrix that optimizes $Q(\mathbf{K}|\mathbf{K}^{(s-1)})$. The pseudo-code is shown in Algorithm \ref{alg:ECM}.

\begin{algorithm}[H]
\label{alg:ECM} 
\caption{Expectation Conditional Maximization (ECM) algorithm}
 \KwIn{ Data $\mathbf{X}$ and an initial precision matrix $\mathbf{K}^{(0)}$.}
 \For{iteration $s=1,\dots,S$}{ 
   E-step: evaluate $Q(\mathbf{K}|\mathbf{K}^{s-1})$ \;
   CM-step: set $\mathbf{K}^{(s)} = \argmax_{\mathbf{K} \in \mathcal{K}} Q(\mathbf{K}|\mathbf{K}^{s-1})$\;
 }
 \KwOut{A sequence $\mathbf{K}^{(0)},\mathbf{K}^{(1)},\dots,\mathbf{K}^{(S)}$ that converges to the MAP estimate.} 
\end{algorithm}
ECM methods in Bayesian structure learning differ in three aspects from each other: their prior, their theoretical properties, and their output. See Table \ref{tab:ECMmethods} for an overview. First, each ECM method uses a different SS prior (see Subsection \ref{sec:MCMCjoint}). Second, each ECM method provides consistent results on the graph and/or the precision matrix. For example, with their ECM-BAGUS method \cite{SSLasso} show that the estimate of the precision matrix $\hat{\mathbf{K}}$ gets arbitrarily close to the true precision matrix $\mathbf{K}^{*}$ as the number of observations goes to infinity. To be specific, under certain assumptions and for a constant $C$, they show that 
\begin{equation}
 ||\hat{\mathbf{K}} - \mathbf{K}^*||_\infty \leq C\sqrt{\frac{\log p }{n}},
 \label{eq:consistent_K_BAGUS}
\end{equation}
where $||\cdot||_\infty$ denotes the $l_{\infty}$ norm. The ECM-Horseshoelike (ECM-HL) method by \cite{HorseshoeLike} proves a similar result. The ECM - BAGUS method is the only ECM method that also provides a consistent result for the graph. Third, ECM methods differ in the output they provide. Like the K-methods of Section \ref{sec:Kmethods}, ECM methods require some post-hoc operation to provide inference on the graph. The ECM-BAGUS method is thereby the only ECM method that outputs the edge inclusion probabilities \eqref{eq:edge_inclusion_prob}. 

A limitation of ECM methods is that they only work for specifically designed priors on the precision matrix. This is overcome by \citet{K_LLA}, who present a method that outputs a MAP estimate of the precision matrix for a large class of priors. 

\begingroup
\setlength{\tabcolsep}{4pt} 
\begin{table}[h]
\small
 \begin{tabular}{l l l l  l l l l l l l}
\hline
Reference &  Name & \multicolumn{3}{c}{$P(k_{ij}|G) $} &   & \multicolumn{2}{c}{Theory} & & \multicolumn{2}{c}{Output}
\\
\cline{3-5} \cline{7-8} \cline{10-11}
& &   $(i,j) \in G$     & $(i,j) \not \in G$ & $i = j$ & &i & ii & & $\mathbf{K}_{\scaleto{MAP}{5pt}}$ & $p_e$   \\
\hline

\cite{SSLasso} & {\small ECM - BAGUS} & DE$(0,v_1)$ & DE$(0,v_0 )$& Exp$(\eta)$ & &\checkmark & \checkmark & & \checkmark & \checkmark \\

{\small \cite{EMGS}} & {\small ECM - EMGS} &  N$(0,v_1)$ & N$(0,v_0)$ &  Exp$(\eta)$ & & &  & & \checkmark &\\

\cite{HorseshoeLike} & {\small ECM - HL} & \multicolumn{2}{c}{Horseshoe-like} & $\propto 1$ & & \checkmark &  & & \checkmark &\\

\hline
\end{tabular}
\caption{\textit{ECM methods for Bayesian structure learning with their corresponding reference, name, prior on $\mathbf{K}$, theoretical properties, and output. DE$(a,b)$ denotes the double exponential distribution with mean $a$ and scale parameter $b$. N$(a,b)$ denotes the normal distribution with mean $a$ and standard deviation $b$. Exp$(\eta)$ is the exponential distribution with rate $\eta$. Note that all marginal priors are not equal to the distributions mentioned in the table due to the positive definite constraint. The theoretical properties are i) the MAP estimate is consistent, and ii) the estimate of the graph is consistent. The output is the MAP estimate of the precision matrix ($\mathbf{K}_{MAP}$) and/or the edge inclusion probabilities $p_e$ of \eqref{eq:edge_inclusion_prob}.  }}
\label{tab:ECMmethods}
\end{table}
\endgroup

\subsection{Regression Approaches}
\label{sec:regression_algorithms}
In Section \ref{sec:frequentistSL}, we saw that structure learning is equivalent to finding the coefficients of $p$ different regressions. That is, estimating the coefficients $\beta_{ij}$ in \eqref{eq:regression} allows for direct estimation of the graphical model, since $\beta_{ij} = \frac{k_{ij}}{k_{ii}}$ and hence $\beta_{ij} = 0 \iff (i,j) \not \in G$. Regression methods use a Bayesian approach to find the coefficients $\beta_{ij}$, thereby offering a simple approach to Bayesian structure learning. The resulting estimates of the graph and the precision matrix, however, come with limited theoretical guarantees.

There are two regression methods: The R - Bayesian Lasso Neighborhood Regression Estimate (R - BLNRE) method by \citet{BLNRE} and the R - Projection (R-P) method by \cite{Projection}. Both put priors on each coefficient $\beta_{ij}$ and create Markov chains $(\beta_{ij}^{(1)},\dots,\beta_{ij}^{(S)})$ for each $\beta_{ij}$. These MCMC chains provide approximate samples from the posterior distribution $P(\beta_{ij}|\mathbf{X})$. Like MCMC-based approaches on the space of precision matrices (Section \ref{sec:Kmethods}), regression methods now use a post-MCMC operation to obtain an estimate of the graph and the precision matrix. 

The R-BLNRE method finds the $\beta_{ij}$ in \eqref{eq:regression} using the Bayesian lasso, a popular algorithm by \citet{parkandcasella}. In the Bayesian lasso the prior on each coefficient $\beta_{ij}$ is set to the double exponential distribution. The mode of the resulting posterior distribution $P(\beta_{ij}|\mathbf{X})$ equals the frequentist estimate obtained by solving the optimization problems in \eqref{eq:BayesianLasso}. 

The R-P method by \cite{Projection} uses the horseshoe prior for the regression coefficients. The same prior is used for the elements of the precision matrix in the K- Horseshoe method by \citet{GraphHorseshoe}. Using so-called projection predictive selection, the samples from the resulting posterior $P(\beta_{ij}|\mathbf{X})$ are used to output an estimate of the graph and precision matrix.

\subsection{Hypothesis Approaches}
Hypothesis methods circumvent the time-consuming exploration of the model space necessary for MCMC-based methods. Instead, they perform inference on the graph directly. To do so, they first formulate, for every $(i,j)$, a null hypothesis $H_0: k_{ij} = 0$ and an alternative hypothesis $H_1: k_{ij} \not = 0$. Then, they calculate the Bayes factor in favour of $H_1$, given by 
\begin{align*}
    BF_{ij} & = \frac{P(\mathbf{X}|H_1)}{P(\mathbf{X}|H_0)}.
\end{align*}
This strategy was introduced by \citet{BayesfactorGiudici}, who derived a closed-form expression of $BF_{ij}$ when the number of observations is larger than the number of variables, \textit{i.e.}, $n>p$. The Hypothesis Leday and Richardson (H-LR) method by \citet{Leday} builds upon this result by deriving a closed-form expression for $BF_{ij}$ that also holds for $p > n$ and show that it is consistent, \textit{i.e.}, $\lim_{n\to \infty}BF_{ij} = 0$ if $H_0$ is true, and $\lim_{n\to \infty}BF_{ij} = \infty$ if $H_1$ is true. The H-LR method scales $BF_{ij}$ to obtain a scale-invariant Bayes factor $sBF_{ij}$. By computing $sBF_{ij}$ for all edges and selecting a threshold $\gamma$, one can obtain a graph estimate $\hat{G}$ by including an edge $(i,j)$ in $\hat{G}$ if $sBF_{ij} > \gamma$. For a prior on the precision matrix, they use the Wishart distribution, which is equivalent to the $G$-Wishart distribution $P(\mathbf{K}|G)$ in Equation \ref{eq:Gwishart} when $G$ is a complete graph. The H-LR method does not use a prior on the graph and is therefore, like the K-methods, not able to incorporate a prior belief on the graphical structure.

The Hypothesis - BGGM (H-BGGM) method by \citet{Williams2020} formulates the hypothesis using the partial correlations $\rho_{ij}$ in \eqref{eq:partialcorr}. The resulting hypotheses are $H_0: \rho_{ij} = 0$, $H_1: \rho_{ij} > 0$, and  $H_2: \rho_{ij} < 0$. They also formulate an unrestricted hypothesis $H_u: \rho_{ij} \in (-1,1)$. These hypotheses allow for the computation of a scale-invariant Bayes factor. The edge exclusion probability $P((i,j) \not \in G)$ can now be calculated as follows:
\begin{align*}
   P((i,j) \not \in G) & = P(H_0 | \mathbf{X}) \\
   & = \frac{P(\mathbf{X}|H_0)P(H_0)}{P(\mathbf{X}|H_0)P(H_0) + P(\mathbf{X}|H_1)P(H_1) + P(\mathbf{X}|H_2)P(H_2)} \\
   & = \frac{BF_{0u}P(H_0)}{BF_{0u}P(H_0) + BF_{1u}P(H_1) + BF_{2u}P(H_2)}.
\end{align*}
Here, $BF_{ku} = \frac{P(\mathbf{X}|H_k)}{P(\mathbf{X}|H_u)} = \frac{P(H_k|\mathbf{X})}{P(H_k)}$ for $k=0,1,2$. $P(H_k)$ is the prior distribution for accepting hypothesis $H_k$. The Bayes factor $BF_{ku}$ is therefore just a ratio of the posterior and prior of the hypotheses $H_k$. Using the Wishart prior for the precision matrix, they show that the prior and posterior partial correlations approximately follow normal distributions. The Bayes factors $BF_{ku}$ can then be computed analytically or by sampling repeatedly from the Wishart distribution and using Equation \ref{eq:partialcorr}. The Wishart distribution, however, is not able to incorporate prior beliefs on the graph. Therefore, they propose two other priors: the Corrected Wishart prior and the matrix-F prior. The H-BGGM method can also be used for confirmatory hypothesis tests. These kinds of hypothesis tests involve more than one partial correlation; for example, $H_0: \rho_{12} > \rho_{34}$ and $H_1: \rho_{12} \leq \rho_{34}$.

\section{Empirical Comparison}
\label{sec:simulation} 

The simulation study in this section contributes to the literature in three ways. First, it includes large-scale instances, \textit{i.e.}, instances with a thousand variables. Until now, Bayesian algorithms are only compared on 250 variables or less, an exception being the K-LRD algorithm by \citet{xChandra2021}. Second, we report information on how long it takes algorithms to achieve good results. This information is missing in Bayesian literature. Third, and most importantly, never before has such a variety of Bayesian algorithms been compared in a single simulation study. All our results can be reproduced by the specified scripts on our GitHub page\footnote{\url{https://github.com/lucasvogels33/Review-paper-Bayesian-Structure-Learning-in-GGMs}}. We discuss the simulation setup in Subsection \ref{sec:simset} and the results in Subsection \ref{sec:sim_results}.

\subsection{Simulation Setup}
\label{sec:simset}

\textbf{Selected algorithms}: 
In Bayesian structure learning literature, new algorithms tend to be compared to frequentist algorithms, or Bayesian algorithms of similar nature, \textit{e.g.}, K-algorithms are compared to K-algorithms, ECM algorithms to ECM algorithms, and so on. We group Bayesian methods into seven categories and show in Table \ref{tab:Bayesian_comparisons} that comparisons between different categories are largely missing in the literature.

\begingroup
\setlength{\tabcolsep}{4pt} 

\begin{table}[ht]
\begin{tabular}{l|ccccccccc}
& \rotatebox{90}{Frequentist} & \rotatebox{90}{RJ}        & \rotatebox{90}{BD} & \rotatebox{90}{SS}        & \rotatebox{90}{G-methods} & \rotatebox{90}{K-methods} & \rotatebox{90}{ECM}       & \rotatebox{90}{Regression} & \rotatebox{90}{Hypothesis} \\
\hline
Frequentist          & \cellcolor{gray} \footnotesize \textcolor{gray}{\textit{missing}}  &  \footnotesize \textcolor{gray}{\textit{missing}}       & \citeyear{BDMCMC1}              & \citeyear{CONCORD} &  \footnotesize \textcolor{gray}{\textit{missing}}   & \citeyear{HorseshoeLike}    & \citeyear{HorseshoeLike} & \citeyear{Projection}  & \citeyear{Williams2020}   \\
RJ                   &             & \cellcolor{gray} \footnotesize \textcolor{gray}{\textit{missing}} &               \citeyear{WWA}         &     \footnotesize \footnotesize \textcolor{gray}{\textit{missing}}        &     \citeyear{Stingo} &     \footnotesize \textcolor{gray}{\textit{missing}}         &     \footnotesize \textcolor{gray}{\textit{missing}}      &    \footnotesize \textcolor{gray}{\textit{missing}}        & \citeyear{Leday}  \\
BD                   &             &           & \cellcolor{gray} \footnotesize \textcolor{gray}{\textit{missing}}              &    \footnotesize \textcolor{gray}{\textit{missing}}        &      \footnotesize \textcolor{gray}{\textit{missing}}         &       \footnotesize \textcolor{gray}{\textit{missing}}        &       \footnotesize \textcolor{gray}{\textit{missing}}     &  \citeyear{Projection} & \citeyear{Williams2020}  \\
SS                   &             &           &                        & \cellcolor{gray} &      \footnotesize \textcolor{gray}{\textit{missing}}        &    \footnotesize \textcolor{gray}{\textit{missing}}          &     \footnotesize \textcolor{gray}{\textit{missing}}      &      \footnotesize \textcolor{gray}{\textit{missing}}      &       \footnotesize \textcolor{gray}{\textit{missing}}     \\
G-methods         &             &           &                        &           & \cellcolor{gray}    &         \footnotesize \textcolor{gray}{\textit{missing}}     &   \footnotesize \textcolor{gray}{\textit{missing}}         &  \footnotesize \textcolor{gray}{\textit{missing}}          &    \footnotesize \textcolor{gray}{\textit{missing}}        \\
K-methods         &             &           &                        &           &              & \cellcolor{gray}    &    \citeyear{HorseshoeLike}       & \citeyear{Projection}           &       \footnotesize \textcolor{gray}{\textit{missing}}        \\
ECM                   &             &           &                        &           &              &              & \cellcolor{gray} &     \footnotesize \textcolor{gray}{\textit{missing}}       &  \footnotesize \textcolor{gray}{\textit{missing}}          \\
Regression           &             &           &                        &           &              &              &           & \cellcolor{gray}  &       \footnotesize \textcolor{gray}{\textit{missing}}     \\
Hypothesis           &             &           &                        &           &              &              &           &            & \cellcolor{gray} 
\end{tabular}
\caption{\textit{Existing and missing empirical comparisons among different classes of Bayesian methods.}}
\label{tab:Bayesian_comparisons}
\end{table}

\endgroup

In this section, we complete most of this table by comparing seven Bayesian algorithms (RJ-WWA, BD-A, SS-O, G-MPLBD, K-Horseshoe, ECM-BAGUS, and H-BGGM) and one frequentist algorithm (glasso). For this simulation study, Bayesian algorithms are chosen based on three criteria: i) the algorithm demonstrates outstanding performance in its category, ii) there is publicly accessible working code for the algorithm, and iii) the algorithm is documented in a published article. No regression algorithm is selected because there is no code available. 
As a prior for the graph we sample the edges from independent and identically distributed Bernoulli distributions. The resulting prior becomes 
\begin{equation}
\label{eq:bernoulli_prior}
P(G) = \delta^{|E|}(1-\delta)^{p(p-1)/2 - |E|},
\end{equation}
where $\delta$ is the prior sparsity of the graph. In literature this prior sparsity is often set to either the uninformative value $\delta=0.5$ \citep{SSLasso} or to a function that decreases with $p$, for example $2/(p-1)$ \citep{WWA}. In this simulation study, we opt for a middle ground and set $\delta=0.2$. Supplementary materials S1 and S2 contain the references and all other settings of the algorithms.

\textbf{Performance metrics}: A metric should measure the distance between an estimate and the truth. In the case of Bayesian structure learning, however, the true posterior distribution $P(G|\mathbf{X})$ is unknown. In the absence of a true posterior distribution, it is therefore common in the literature to opt for the second best choice: compare the estimate of the posterior against the true graph $G^*$. For lack of a better alternative, we adopt the same strategy too. 

We compare the selected algorithms based on the accuracy of the edge inclusion probabilities $p_e$ defined in \eqref{eq:edge_inclusion_prob}. Let $P$ be the matrix with elements $p_e$. We refer to $P$ as the edge inclusion matrix. We use three metrics for the accuracy of the edge inclusion probabilities: AUC, $Pr^+$, and $Pr^-$. The AUC represents the area under the Receiver Operating Characteristic (ROC) curve. It reflects to what degree the edge inclusion probabilities of links in the true graph are higher than those of links not in the true graph. It ranges from $0$ (worst) to $1$ (best). An AUC of $0.5$ means the edge inclusion matrix is as good as a random guess. The glasso and K-Horseshoe algorithms do not output edge inclusion probabilities. To calculate the AUC of the glasso algorithm, we solve the minimization problem in \eqref{eq:glasso} for different values of $\lambda$. The area under the resulting ROC curve gives us the AUC. For the K-Horseshoe algorithm we use the same strategy as \cite{GraphHorseshoe}, \textit{i.e.}, the ROC curve is generated by varying the length of posterior credible intervals from 1\% to 99\%. 

The AUC is a measure for the ranking of the edge inclusion probabilities but does not convey their magnitude. We therefore add two metrics for the magnitude: one measuring the algorithm's ability to predict the presence of an edge ($Pr^+$), the other the ability to predict the absence of an edge ($Pr^-$). Formally, we have
\begin{equation}
 Pr^+ = \frac{1}{|E^*|}\sum_{(i,j) \in G^*}p_e
 \label{eq:pplus}
\end{equation}
and
\begin{equation}
 Pr^- = \frac{1}{|E^{*-}|}\sum_{(i,j) \not \in G^*}p_e,
 \label{eq:pmin}
\end{equation}
where $|A|$ denotes the number of elements in the set $A$ and $E^{*-}$ denotes the set of links that are not in the true graph. $Pr^+$ ranges from $0$ (worst) to $1$ (best), whereas $Pr^-$ ranges from $1$ (worst) to $0$ (best). As before, these metrics need to be redefined for the glasso and K-Horseshoe algorithms, since these algorithms do not output edge inclusion probabilities, but a point estimate of the true graph. By setting $p_e = 1$ if $e \in \hat{G}$, and zero if $e \not \in \hat{G}$, we can still calculate $Pr^+$ and $Pr^-$. Note that $Pr^+$ ($Pr^-$) then becomes the true (false) positive rate of the estimate $\hat{G}$. Concerns might be raised about the fairness when comparing the $Pr^+$ and $Pr^-$ metrics from a point estimate $\hat{G}$ with those from the posterior edge inclusion probabilities $p_e$. However, in both cases, the $Pr^+$ and $Pr^-$ metrics do provide valuable information and are thus included. We do recommend the reader proceed with caution when directly comparing the $Pr^+$ and $Pr^-$ metrics from the glasso and K-horseshoe method with those from other algorithms. 

We also measure the computational cost of each algorithm. We first run each Bayesian algorithm until visual inspection shows convergence and denote with $P$ the final edge inclusion matrix. The computational cost $T$ of an algorithm is then defined as the time it takes the algorithm to produce an AUC within 0.01 of its final AUC. That is,
\begin{equation}
\label{eq:AUC_convergence}
T = \min(t:|AUC(P_t) - AUC(P)| < \epsilon), 
\end{equation}
where we set $\epsilon = 0.01$. Here, $P_t$ denotes the edge inclusion matrix after $t$ seconds of running the algorithm and $AUC(P)$ denotes the AUC corresponding to an edge inclusion matrix $P$. In real-world applications, the true graph is unknown and the computational cost $T$ is not defined. For convergence assessments without knowing the true graph, we refer to Section \ref{sec:application}. The computational cost of the H-BGGM and glasso algorithms can not be defined using \eqref{eq:AUC_convergence}. Instead, we define their computation cost $T$ as the total time it takes these algorithms to provide their edge inclusion probabilities. We note that both the glasso and ECM-BAGUS algorithms require parameter tuning. We include this time in the computational cost.

\textbf{Simulation instances}: We include small ($p=10)$, medium ($p=100$) and large ($p=1000$) scale graphs with a low $(n=20\log p)$ and high $(n=350\log p)$ number of observations. The use of the logarithm here is motivated by the performance bound given by \eqref{eq:consistent_K_BAGUS}. We consider the following two graph types:
\begin{enumerate}
   \item \textsf{Random}: a graph in which every edge $(i,j)$ is drawn from an independent Bernoulli distribution with probability $\delta$. 
   \item \textsf{Cluster}: a graph with two clusters for $p \in \{10,100\}$ and eight clusters for $p=1000$. Each cluster has the same structure as the  \textsf{Random} graph. 
\end{enumerate}
The value of $\delta$ is chosen so that the resulting edge density in the graphs is $10\%$ for $p=10$, $1\%$ and $10\%$ for $p=100$, and $0.1\%$ for $p=1000$. For every generated graph $G^*$, the corresponding precision matrix $\mathbf{K}^*$ is then sampled from the $G$-Wishart distribution. That is $\mathbf{K}^* \sim W_{G^*}(3, \mathbf{I})$. For each generated $G^*$ and $\mathbf{K}^*$, we then sample $n$ observations from the $p-$variate normal distribution with covariance matrix $\mathbf{\Sigma}^* = \mathbf{K}^{*{-1}}$ and mean zero. For $p \in \{10,100,1000\}$, each $n$, each graph type, and each density, we repeat the process 16 times which is generally enough to reach low ($<0.025$) standard errors on AUC, $Pr^+$ and $Pr^-$.  

\subsection{Results}
\label{sec:sim_results}
\begin{figure}[H]
     \centering
     \begin{subfigure}[b]{.32\textwidth}
         \centering
        \includegraphics[width=\linewidth]{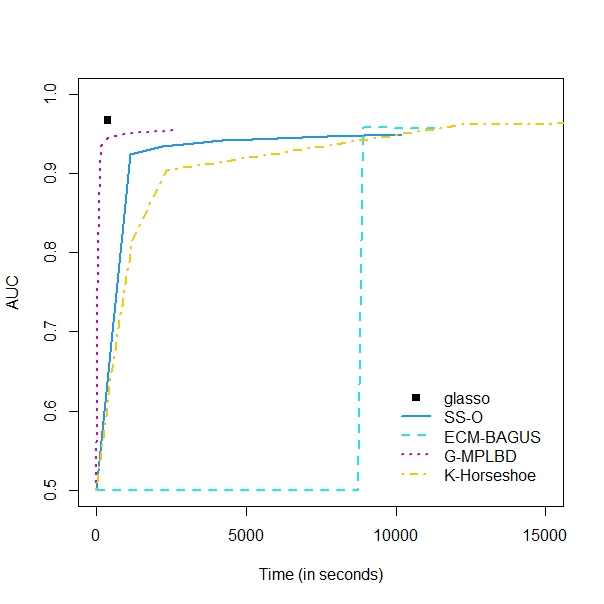}
     \end{subfigure}
     \begin{subfigure}[b]{.32\textwidth}
         \centering
        \includegraphics[width=\linewidth]{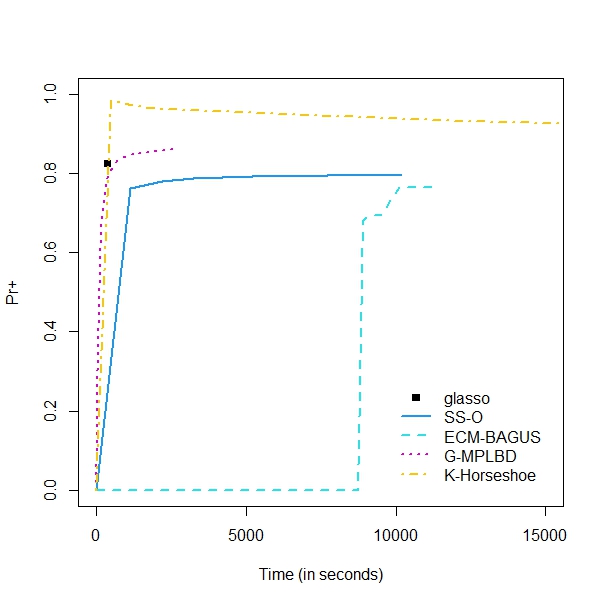}
     \end{subfigure}
     \begin{subfigure}[b]{.32\textwidth}
         \centering
        \includegraphics[width=\linewidth]{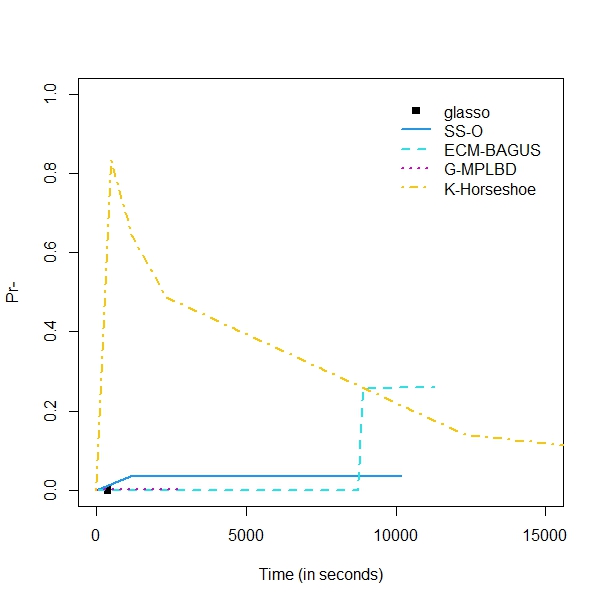}
     \end{subfigure}
     \hfill
    \caption{\textit{AUC (left), $Pr^+$ (middle) and $Pr^-$ (right) over time. The figures show averages over 16 replications on an instance with $p=1000$, $n=1050$ on the  \textsf{Random} graph with density $0.1\%$.     }}
    \label{fig:p1000}
\end{figure}

The results on the computational costs $T$, AUC, $Pr^-$ and $Pr^+$ are presented, respectively, in Tables \ref{table:T}, \ref{table:AUC}, \ref{table:pmin}, and \ref{table:pplus}. If there is one conclusion to be drawn from these results, then it is this: the idea that Bayesian algorithms lack the computational advantage of their frequentist counterparts is outdated. Whether with 10, 100, or 1000 variables, one or more Bayesian algorithms can estimate the full posterior distribution or any function thereof at the same time that the glasso algorithm can provide a point estimate. Moreover, there is no evidence of the glasso algorithm outperforming Bayesian algorithms on AUC, $Pr^+$, or $Pr^-$ either. It underlines the staggering progress Bayesian algorithms have made over the last decade. This progress is illustrated too in Figure \ref{fig:p1000}, which shows the convergence of performance metrics for each algorithm on a large scale instance $(p=1000)$: the AUC, $Pr^+$ and $Pr^-$ values of the G-MPLBD, SS-O, and K-Horseshoe algorithms converge in approximately the same time as the glasso algorithm.

Now, let us look into the results in more detail, starting with the computational cost in Table \ref{table:T}. Unsurprisingly, the glasso algorithm is fast. It provides a point estimate of the graph within seconds for small and medium-sized problems and within 10 minutes for large-scale problems. The majority of these 10 minutes is spent on tuning the regularization parameter $\lambda$ in \eqref{eq:glasso}. More surprising is the outstanding performance of the G-MPLBD algorithm that allows inference on the entire posterior within the same time as the glasso provides a point estimate. The most time-consuming algorithms are the RJ-WWA and BD-A algorithms, which are slow to run on large-scale instances. This is due to the computationally expensive sampling from the $G$-Wishart distribution at every iteration and the slow edge-by-edge exploration of the state space. Moreover, both algorithms benefit from parallel computing but are run on one core in this simulation study. The H-BGGM algorithm gives errors for the high dimensional case $n < p$ and for the case $p=1000$. The fields corresponding to these cases are therefore left empty.

In terms of the AUC (Table \ref{table:AUC}), most algorithms perform similarly. Noteworthy is the overall underperformance of the ECM-BAGUS algorithm and the underperformance of the glasso algorithm in medium-sized problems ($p=100$) with a high edge density ($10$\%). The $Pr^-$ values (Table \ref{table:pmin}) are comparable across most algorithms too with estimates close to the desired value of zero. Only the ECM-BAGUS and BGGM algorithms perform significantly worse, the latter even assigning in some cases an average edge inclusion probability of $0.88$ to edges that are not in the true graph. Looking at the $Pr^+$ values in Table \ref{table:pplus}, we see more variability between algorithms. For instances with a high number of observations, the K-Horseshoe algorithm performs best. For large-scale graphs, the K-horseshoe algorithm outperforms the others too. The ECM-BAGUS algorithm performs slightly worse than others in most instances. The $Pr^+$ values of the BGGM algorithm seem close to optimal but are only marginally more than their $Pr^-$ values.

The results also reveal that it is not so much the structure of the graph (\textsf{Random} or  \textsf{Cluster}) as the density that influences the performance of algorithms. As expected, more observations lead to better results. Generally, $n = 350 \log p$ observations are sufficient to yield good results.

\spacingset{1.5}
\begin{table}[h]
\centering
\begin{tabular}{l l l l l l l l l l l l}
\arrayrulecolor{blue} \\ \\
$p$ & Graph & Density & $n$ &  
\begin{rotate}{45} glasso \end{rotate} & 
\begin{rotate}{45} RJ-WWA \end{rotate} &
\begin{rotate}{45} BD-A \end{rotate}& 
\begin{rotate}{45} SS-O \end{rotate}& 
\begin{rotate}{45} \small ECM-BAGUS \end{rotate}& 
\begin{rotate}{45} \small G-MPLBD \end{rotate}& 
\begin{rotate}{45} \small K-Horseshoe \end{rotate}& 
\begin{rotate}{45} H-BGGM \end{rotate}\\
\toprule
10 &  \textsf{Random}  &  10 \% & 20  &2  & 1  & 0  & 0  & 1  & 0  & 0  & 0   \\ 
 &   &  10 \% & 350  &2  & 0  & 0  & 0  & 1  & 0  & 0  & 0   \\ 
 \cline{2-12}
 &  \textsf{Cluster}  &  10 \% & 20  &2  & 1  & 0  & 0  & 1  & 0  & 0  & 0   \\ 
 &   &  10 \% & 350  &2  & 0  & 0  & 0  & 1  & 0  & 0  & 0   \\ 
\hline
100 &  \textsf{Random}  &  1 \% & 40  &3  & 3623  & 5313  & 5  & 84  & 4  & 49  & -\\ 
 &    &  1 \% & 700  &3  & 147  & 1835  & 19  & 76  & 2  & 3  & 11   \\ 
 &    &  10 \% & 40  &5  & 2794  & 9018  & 10  & 252  & 6  & 32  & -\\ 
 &    &  10 \% & 700  &5  & 746  & 6093  & 12  & 482  & 8  & 18  & 10   \\ 
 \cline{2-12}
 &  \textsf{Cluster}  &  1 \% & 40  &3  & 620  & 4496  & 13  & 83  & 4  & 11  & -\\ 
 &    &  1 \% & 700  &3  & 159  & 3013  & 19  & 70  & 1  & 5  & 11   \\ 
 &    &  10 \% & 40  &4  & 3489  & 10368  & 13  & 186  & 5  & 17  & -\\ 
 &    &  10 \% & 700  &3  & 1569  & 6465  & 10  & 349  & 6  & 10  & 11   \\ 
 \hline
1000 &  \textsf{Random}  &  0.1 \% & 60  &606  & -&-&9063  & 15057  & 624  & 24064  & -\\ 
 &    &  0.1 \% & 1050  &354  & -&-&4007  & 8907  & 492  & 12300  & -\\ 
 \cline{2-12}
 &  \textsf{Cluster}  &  0.1 \% & 60  &672  & -&-&8510  & 15710  & 623  & 7451  & -\\ 
 &    &  0.1 \% & 1050  &366  & -&-&3823  & 8224  & 481  & 12389  & -\\ 

\bottomrule
\end{tabular}
\caption{\textit{Computational cost $T$ \eqref{eq:AUC_convergence} in seconds of the algorithms for different instances. The values are averages over 16 replications. The ``-" entry indicates that an algorithm did not produce results on that instance.}}
\label{table:T}
\end{table}

\begin{table}[H]
\centering
\begin{tabular}{l l l l l l l l l l l l}
\arrayrulecolor{blue} \\ \\
$p$ & Graph & Density & $n$ &   
\begin{rotate}{45} glasso \end{rotate} & 
\begin{rotate}{45} RJ-WWA \end{rotate} &
\begin{rotate}{45} BD-A \end{rotate}& 
\begin{rotate}{45} SS-O \end{rotate}& 
\begin{rotate}{45} \small ECM-BAGUS \end{rotate}& 
\begin{rotate}{45} \small G-MPLBD \end{rotate}& 
\begin{rotate}{45} \small K-Horseshoe \end{rotate}& 
\begin{rotate}{45} H-BGGM \end{rotate}\\
\toprule
10 &  \textsf{Random}  &  10 \% & 20  &0.7  & 0.72  & 0.73  & 0.73  & 0.7  & 0.72  & 0.7  & 0.67  \\ 
 &    &  10 \% & 350  &0.94  & 0.95  & 0.95  & 0.93  & 0.91  & 0.95  & 0.94  & 0.93  \\ 
 \cline{2-12}
 &  \textsf{Cluster}  &  10 \% & 20  &0.83  & 0.82  & 0.82  & 0.81  & 0.79  & 0.82  & 0.81  & 0.75  \\ 
 &    &  10 \% & 350  &0.93  & 0.93  & 0.93  & 0.91  & 0.92  & 0.94  & 0.92  & 0.92  \\ 
\hline
100 &  \textsf{Random}  &  1 \% & 40  &0.83  & 0.83  & 0.81  & 0.84  & 0.76  & 0.81  & 0.84  & -\\ 
 &    &  1 \% & 700  &0.96  & 0.96  & 0.96  & 0.94  & 0.94  & 0.95  & 0.96  & 0.95  \\ 
 &    &  10 \% & 40  &0.67  & 0.7  & 0.68  & 0.71  & 0.59  & 0.67  & 0.71  & -\\ 
 &    &  10 \% & 700  &0.78  & 0.92  & 0.91  & 0.89  & 0.8  & 0.9  & 0.92  & 0.87  \\ 
 \cline{2-12}
 &  \textsf{Cluster}  &  1 \% & 40  &0.83  & 0.85  & 0.83  & 0.83  & 0.78  & 0.81  & 0.85  & -\\ 
 &    &  1 \% & 700  &0.95  & 0.96  & 0.96  & 0.94  & 0.94  & 0.94  & 0.95  & 0.94  \\ 
 &    &  10 \% & 40  &0.71  & 0.77  & 0.72  & 0.75  & 0.61  & 0.7  & 0.73  & -\\ 
 &    &  10 \% & 700  &0.84  & 0.94  & 0.93  & 0.91  & 0.84  & 0.92  & 0.93  & 0.87  \\ 
\hline
1000 &  \textsf{Random}  &  0.1 \% & 60  &0.86  & -&-&0.84  & 0.84  & 0.72  & 0.84  & -\\ 
 &    &  0.1 \% & 1050  &0.97  & -&-&0.95  & 0.96  & 0.95  & 0.96  & -\\ 
 \cline{2-12}
 &  \textsf{Cluster}  &  0.1 \% & 60  &0.84  & -&-&0.82  & 0.84  & 0.7  & 0.84  & -\\ 
 &    &  0.1 \% & 1050  &0.96  & -&-&0.95  & 0.95  & 0.95  & 0.96  & -\\ 

\bottomrule
\end{tabular}
\caption{\textit{AUC scores of the algorithms for different instances. The AUC reaches its best score at $1$ and its worst at $0$. The values are averages over 16 replications. The ``-" entry indicates that an algorithm did not produce results on that instance.}}
\label{table:AUC}
\end{table}

\begin{table}[h]
\centering
\begin{tabular}{l l l l l l l l l l l l}
\arrayrulecolor{blue} \\ \\
$p$ & Graph & Density & $n$ &   
\begin{rotate}{45} glasso \end{rotate} & 
\begin{rotate}{45} RJ-WWA \end{rotate} &
\begin{rotate}{45} BD-A \end{rotate}& 
\begin{rotate}{45} SS-O \end{rotate}& 
\begin{rotate}{45} \small ECM-BAGUS \end{rotate}& 
\begin{rotate}{45} \small G-MPLBD \end{rotate}& 
\begin{rotate}{45} \small K-Horseshoe \end{rotate}& 
\begin{rotate}{45} H-BGGM \end{rotate}\\
\hline
10 &  \textsf{Random}  &  10 \% & 20  &0.05  & 0.09  & 0.1  & 0.08  & 0.18  & 0.08  & 0.06  & 0.88  \\ 
&    &  10 \% & 350  &0.05  & 0.02  & 0.03  & 0.03  & 0.15  & 0.01  & 0.09  & 0.33  \\ 
\cline{2-12}
&  \textsf{Cluster}  &  10 \% & 20  &0.04  & 0.08  & 0.09  & 0.08  & 0.2  & 0.07  & 0.02  & 0.88  \\ 
&    &  10 \% & 350  &0.06  & 0.02  & 0.03  & 0.03  & 0.12  & 0.01  & 0.07  & 0.31  \\ 
\hline
100 &  \textsf{Random}  &  1 \% & 40  &0  & 0.05  & 0.07  & 0.09  & 0.16  & 0.03  & 0  & -\\ 
&    &  1 \% & 700  &0.01  & 0.01  & 0.02  & 0.03  & 0.11  & 0.01  & 0  & 0.23  \\ 
&    &  10 \% & 40  &0.13  & 0.1  & 0.12  & 0.12  & 0.12  & 0.04  & 0.01  & -\\ 
&    &  10 \% & 700  &0.17  & 0.03  & 0.04  & 0.05  & 0.15  & 0  & 0.04  & 0.23  \\ 
\cline{2-12}
&  \textsf{Cluster}  &  1 \% & 40  &0  & 0.05  & 0.07  & 0.09  & 0.18  & 0.03  & 0  & -\\ 
&    &  1 \% & 700  &0  & 0.01  & 0.02  & 0.03  & 0.14  & 0.01  & 0  & 0.23  \\ 
&    &  10 \% & 40  &0.06  & 0.08  & 0.11  & 0.11  & 0.12  & 0.04  & 0.01  & -\\ 
&    &  10 \% & 700  &0.11  & 0.02  & 0.03  & 0.05  & 0.12  & 0  & 0.03  & 0.23  \\ 
\hline
1000 &  \textsf{Random}  &  0.1 \% & 60  &0  & -&-&0.18  & 0.29  & 0.1  & 0.05  & -\\ 
&    &  0.1 \% & 1050  &0  & -&-&0.04  & 0.26  & 0  & 0.05  & -\\ 
\cline{2-12}
&  \textsf{Cluster}  &  0.1 \% & 60  &0  & -&-&0.18  & 0.32  & 0.1  & 0.05  & -\\ 
&    &  0.1 \% & 1050  &0  & -&-&0.04  & 0.23  & 0  & 0.05  & -\\ 

\bottomrule
\end{tabular}
\caption{\textit{$Pr^-$ scores \eqref{eq:pmin} of the algorithms for different instances. The $Pr^-$ reaches its best score at $0$ and its worst at $1$. The values are averages over 16 replications. The ``-" entry indicates that an algorithm did not produce results on that instance.}}
\label{table:pmin}
\end{table}

\begin{table}[H]
\centering
\begin{tabular}{l l l l l l l l l l l l}
\arrayrulecolor{blue} \\ \\
$p$ & Graph & Density & $n$ &   
\begin{rotate}{45} glasso \end{rotate} & 
\begin{rotate}{45} RJ-WWA \end{rotate} &
\begin{rotate}{45} BD-A \end{rotate}& 
\begin{rotate}{45} SS-O \end{rotate}& 
\begin{rotate}{45} \small ECM-BAGUS \end{rotate}& 
\begin{rotate}{45} \small G-MPLBD \end{rotate}& 
\begin{rotate}{45} \small K-Horseshoe \end{rotate}& 
\begin{rotate}{45} H-BGGM \end{rotate}\\
\hline
10 &  \textsf{Random}  &  10 \% & 20  &0.38  & 0.38  & 0.4  & 0.35  & 0.31  & 0.4  & 0.39  & 0.9  \\ 
&    &  10 \% & 350  &0.87  & 0.77  & 0.78  & 0.56  & 0.69  & 0.81  & 0.85  & 0.9  \\ 
\cline{2-12}
&  \textsf{Cluster}  &  10 \% & 20  &0.45  & 0.45  & 0.47  & 0.4  & 0.35  & 0.5  & 0.45  & 0.91  \\ 
&    &  10 \% & 350  &0.87  & 0.79  & 0.8  & 0.58  & 0.73  & 0.82  & 0.88  & 0.89  \\ 
\hline
100 &  \textsf{Random}  &  1 \% & 40  &0.3  & 0.46  & 0.5  & 0.5  & 0.46  & 0.47  & 0.41  & -\\ 
&    &  1 \% & 700  &0.83  & 0.84  & 0.84  & 0.63  & 0.82  & 0.84  & 0.87  & 0.9  \\ 
&    &  10 \% & 40  &0.3  & 0.31  & 0.32  & 0.29  & 0.22  & 0.22  & 0.24  & -\\ 
&    &  10 \% & 700  &0.59  & 0.73  & 0.72  & 0.57  & 0.58  & 0.68  & 0.79  & 0.76  \\ 
\cline{2-12}
&  \textsf{Cluster}  &  1 \% & 40  &0.26  & 0.56  & 0.51  & 0.5  & 0.46  & 0.47  & 0.43  & -\\ 
&    &  1 \% & 700  &0.81  & 0.83  & 0.84  & 0.61  & 0.83  & 0.84  & 0.81  & 0.9  \\ 
&    &  10 \% & 40  &0.27  & 0.33  & 0.35  & 0.32  & 0.24  & 0.26  & 0.26  & -\\ 
&    &  10 \% & 700  &0.59  & 0.73  & 0.73  & 0.59  & 0.63  & 0.68  & 0.8  & 0.77  \\ 
\hline
1000 &  \textsf{Random}  &  0.1 \% & 60  &0.24  & -&-&0.61  & 0.52  & 0.38  & 0.62  & -\\ 
&    &  0.1 \% & 1050  &0.83  & -&-&0.8  & 0.76  & 0.86  & 0.91  & -\\ 
\cline{2-12}
&  \textsf{Cluster}  &  0.1 \% & 60  &0.2  & -&-&0.59  & 0.5  & 0.36  & 0.62  & -\\ 
&    &  0.1 \% & 1050  &0.8  & -&-&0.78  & 0.81& 0.85  & 0.91  & -\\ 
\bottomrule
\end{tabular}
\caption{\textit{$Pr^+$ scores \eqref{eq:pplus} of the algorithms for different instances. The $Pr^+$ reaches its best score at $1$ and its worst at $0$. The values are averages over 16 replications. The ``-" entry indicates that an algorithm did not produce results on that instance.}}
\label{table:pplus}
\end{table}

\spacingset{1.9}

\section{Case Study: Human Gene Expression}
\label{sec:application}
In this section we demonstrate the capabilities of Bayesian structure learning on a real-world data set. We discuss the challenges that arise when applying Bayesian approaches in practice and present ways to tackle these challenges. Consequently, this section serves as a guideline for practitioners who are new to the field.

We use the SS-O algorithm and the BD-A algorithm to perform inference on a gene network between $p=100$ genes using genetic data of $n=60$ unrelated individuals. Several Bayesian structure learning methods are analyzed on this dataset; see, for example, \citep{BDMCMC1,GraphHorseshoe,WWA,MPLpaper}. This dataset can therefore be seen as a benchmark for Bayesian structure learning. We refer to \citet{Stranger2007} and \citet{Bhadra2013} for more information on the collection of the data. 

Genes are specific sequences of DNA. Gene expression is the process by which a gene produces a protein that influences the functioning of an organism. Some of these proteins directly influence the organism by, for example, initializing a process to break down food. Other proteins, however, only serve to ``activate'' other genes, which in turn make proteins that activate other genes, and so on. These activation relationships can be shown in a gene network, where every node corresponds to a gene and every edge indicates an activation relationship between two genes. Discovering these gene networks plays a key role in understanding disease susceptibility and therefore, ultimately, in treatment and public health \citep{Stranger2007}. We obtain a sample $\mathbf{X}$ of our variables $X_i$, $i=1,\dots,p$, where $\mathbf{X}$ is an $n \times p$ matrix with elements $x_{li}$ denoting the level of protein corresponding to gene $i=1,\dots,p$ measured in individual $l=1,\dots,n$.

\begin{figure}[H]
     \centering
     \includegraphics[width=\linewidth,height=125pt]{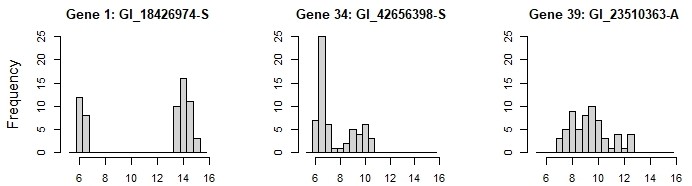}
    \caption{\textit{Univariate histograms of three randomly selected genes in the human gene expression dataset.}}
    \label{fig:gene_histograms}
\end{figure}

Structure learning depends on the assumption that the data comes from a multivariate normal distribution with a mean zero. However, in the gene data set, the univariate histograms of each gene are clearly not normally distributed. Neither are their means zero, see Figure \ref{fig:gene_histograms}. In fact, the normality and zero mean assumptions are rarely met in practice. Fortunately, one can transform our non-Gaussian continuous variables $X_1,\dots, X_p$ to Gaussian variables $Z := (Z_1,\dots, Z_p)$ with mean zero, \textit{i.e.}, $Z \sim \mathcal{N}(\mathbf{0},\mathbf{\Sigma}_Z)$, such that the sparsity of $\mathbf{K}_Z = \mathbf{\Sigma}_Z^{-1}$ still encodes the conditional (in)dependence between the variables $X_1,\dots, X_p$ as in \eqref{eq:relationKandG}; see, for example \cite{Liu2009, Liu2012}. The \textit{huge.npn}() function in the \textit{Huge} R package and the \textit{bdgraph.npn}() function in the \textit{BDgraph} R package perform this transformation efficiently. One can now apply Bayesian structure learning algorithms on the transformed dataset $\mathbf{Z}$. These transformations are only possible for continuous data. For discrete or binary data, Gaussian copula graphical models can be used to perform structure learning; see Section \ref{sec:beyond_Gaussian} for an overview.
We choose to work with MCMC algorithms to perform inference on the posterior of the gene network. These algorithms offer a wide variety of output and are computationally feasible on this 100 node graph. Specifically, we use the SS-O algorithm and the BD-A algorithm, but the same steps apply to other MCMC algorithms. 

First, we select the tuning parameters. For the SS-O algorithm, we follow \cite{SSWang} and select $v_0 =0.02$, $v_1=2$ and $\eta=1$. One could also opt for the more time-consuming approach and run the algorithm several times for different parameters, and select the parameters that perform best according to some criterion \citep{SSLasso}. The BD-A algorithm has no tuning parameters. 

Second, we need to select a prior for the graph. Generally, the prior is selected using prior knowledge about the sparsity of the true graph. Here, we select the prior given by  \eqref{eq:bernoulli_prior}, where we select the prior sparsity equal to $\delta = 0.2$, in line with previous studies on the gene expression dataset; see, for example, \cite{WWA} and \cite{GraphHorseshoe}. 

\begin{figure}[h]
     \centering
     \includegraphics[width=1\linewidth]{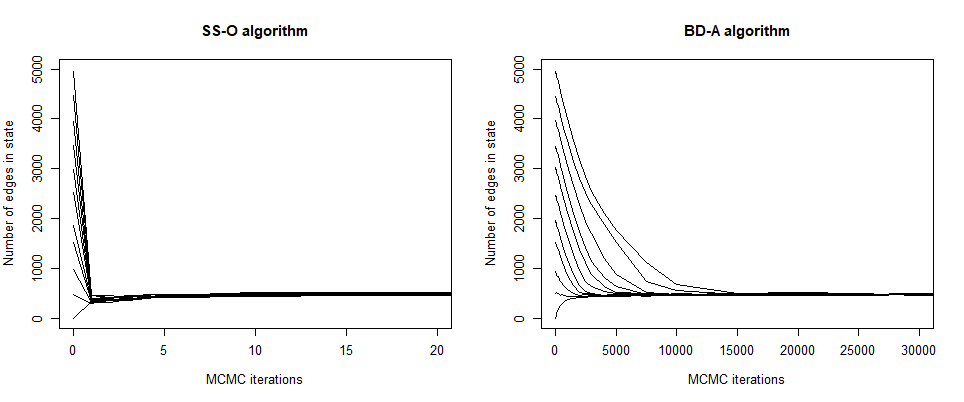}
    \caption{\textit{Number of edges in the visited graph per MCMC iteration for 10 replications. The density of the initial graph differs per replication, ranging from $0\%$ (empty graph) to $100\%$ (full graph).}}
    \label{fig:edge_convergence}
\end{figure}

Now, the correct MCMC settings need to be selected. That is, the number of burn-in iterations, the number of MCMC iterations, and the initial graph $G^{(0)}$ of the Markov Chain. The initial graph should not be important, since the Markov chain should converge to the posterior distribution regardless of the starting graph. To confirm that this indeed happens, we run the algorithms several times, each time from a different initial graph, and report the number of edges in each Markov chain state. The resulting plots (Figure \ref{fig:edge_convergence}) indeed show that, regardless of the initial graph, the Markov chain converges to the same neighbourhood. Moreover, we can deduce that approximately the first 10 (SS-O algorithm) and the first 15,000 iterations (BD-A algorithm) can be discarded as burn-in iterations. To be on the safe side, we select burn-in periods of 100 (SS-O) and 25,000 (BD-A) iterations. 

\begin{figure}[ht]
     \centering
     \includegraphics[width=1\linewidth]{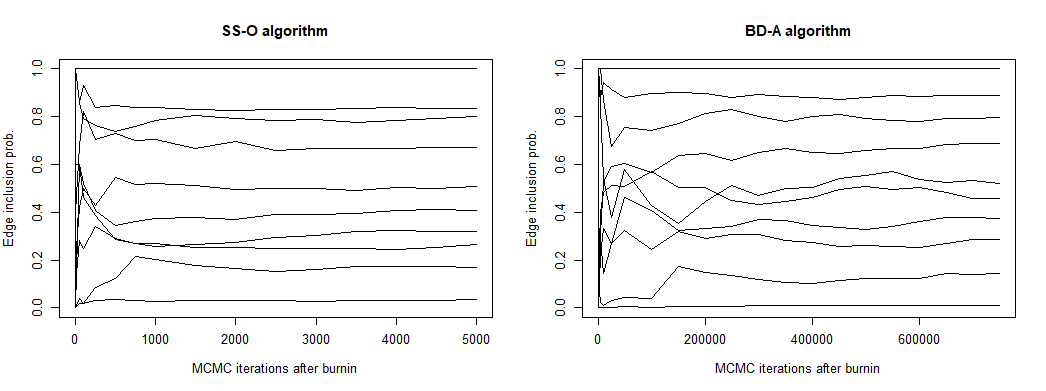}
    \caption{\textit{The edge inclusion probabilities of ten randomly selected edges for every state in the Markov chain.}}
    \label{fig:link_convergence}
\end{figure}

To select the number of MCMC iterations (after burn-in), we plot the edge inclusion probabilities of ten randomly selected links at every state. The resulting plots (Figure \ref{fig:link_convergence}) show that convergence happens in less iterations for the SS-O algorithm (approximately 5,000) than for the BD-A algorithm (approximately 750,000). This is due to the good mixing of the SS-O algorithm versus the slow edge-by-edge exploration of the BD-A algorithm. We set the amount of MCMC iterations accordingly: 5,000 for the SS-O algorithm and 750,000 for the BD-A algorithm. 

\begin{figure}[h]
     \centering
     \includegraphics[width=1\linewidth]{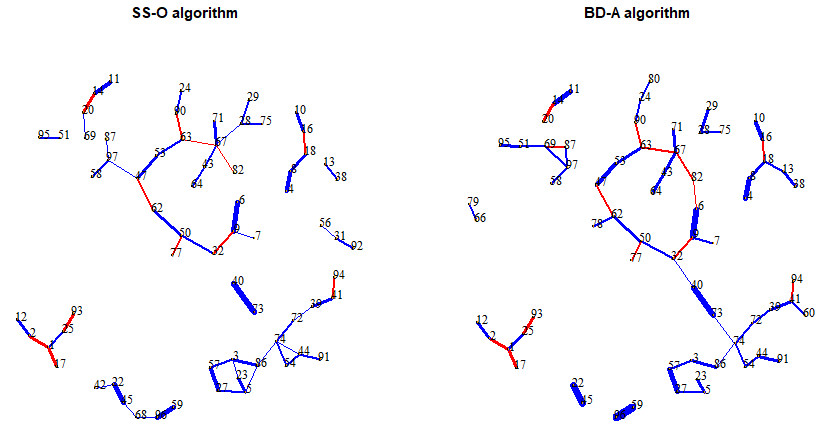}
    \caption{\textit{Gene networks inferred by the SS-O algorithm (left) and the BD-A algorithm (right). Only the 60 edges with the highest edge inclusion probabilities are shown. The width of the edge indicates the strength of the partial correlation. The color represents the sign of the partial correlation: red (negative) and blue (positive). See supplementary material S5 for the gene names corresponding to the node numbering.}}
    \label{fig:gene_network}
\end{figure}

Figure \ref{fig:gene_network} shows the gene networks inferred by the BD-A algorithm and the SS-O algorithm. The networks show similarities in the both the sign (positive or negative) and the size of the partial correlations. Moreover, the networks show similarities with the existing literature. The five most connected genes of the K-Horseshoe algorithm \citep{GraphHorseshoe} all appear among the most connected genes of our analysis. Moreover, most tree structures of Figure \ref{fig:gene_network} appear exactly in the recovered network given by \citet[Figure 4]{Bhadra2013}. 

The impact of the prior density on the posterior remains relatively unexplored in Bayesian structure learning literature. In order to test the sensitivity of the algorithms to the priors, we ran the algorithms with a sparse ($1\%$ density), average ($20\%$ density) and dense ($50\%$ density) graph prior. We compare the boxplots of the resulting edge inclusion probabilities in Figure S1 (SS-O algorithm) and in Figure S2 (BD-A algorithm) in Section S4 in the supplementary materials. As expected, a higher prior density results in higher edge inclusion probabilities with the SS-O algorithm being slightly more sensitive to a higher prior density. For all prior densities, however, both algorithms are able to distinguish between edges with a low and high edge inclusion probability.

\section{Conclusion and Future Research}
\label{sec:discussion}

In structure learning, Bayesian methods constitute a powerful alternative to frequentist ones by providing inference on the entire posterior. The argument that the speed and simplicity of frequentist methods are superior is waning as Bayesian methods have improved significantly in both aspects. Bayesian methods that provide accurate solutions to thousand variable problems within mere minutes are now a reality and easily accessible in software packages. This section starts with a short overview of structure learning literature beyond the basic Gaussian case and ends with suggestions for further research.

\subsection{Extension Beyond Basic GGMs}
\label{sec:beyond_Gaussian}

While the majority of research on graphical modeling has concentrated on Gaussian graphical models, significant efforts have been directed toward extending beyond basic GGMs over the past decade. We provide a brief overview of these extensions here. For comprehensive reviews, works such as \cite{Liang2023, handbookofgraphicalmodels} offer detailed insights. Additionally, these developments merit a dedicated review paper, highlighting their importance in the field.

\textbf{Beyond Gaussianity assumption}:
Graphical modeling for non-Gaussian datasets, including types like single nucleotide polymorphisms, RNA sequencing, and household income data, is increasingly relevant in data science. Over the past decade, adapting models to handle such non-Gaussian data has been a major research area. For continuous non-Gaussian datasets, like the gene interaction data discussed in Section \ref{sec:application}, a two-step method involving a non-parametric data-Gaussianized transformation \citep{Liu2009} followed by GGM application is effective. This method also extends to multivariate discrete data, as shown in \citet{jia2017learning} and \citet[Chapter 5]{Liang2023}. For multivariate discrete or count data, Gaussian copula graphical models offer an alternative, as detailed in \cite{vinciotti2022bayesian, humanmobility, dobra2011copula}. For mixed data types, Bayesian structure learning methods like those in \cite{dobra2011copula, duputryen}, that use a latent GGM with a copula approach \cite{hoff2007extending}, are suitable. 


\textbf{Multiple Gaussian graphical models}:
Multiple GGMs, as an extension to standard GGMs, are useful for jointly analyzing data from multiple sources (also known as heterogeneous data), \textit{e.g.}, neurological data measured at multiple timescales, or joint neurological, genetic, and phenotypic data. For these types of data, \cite{peterson2015bayesian} use Markov random field prior \citep{li2010bayesian} to model a super-graph linking different graphical models. \cite{SMCTan} uses a logistic regression model to link the connectivity of nodes to covariates specific to each graph. More recently, \cite{li2019bayesian} developed doubly spike-and-slab mixture priors as a new class of priors for joint estimation of multiple graphical models.

\textbf{Coloured Gaussian graphical models}: 
These are a class of GGMs with additional symmetry restrictions in the form of equality constraints on the parameters \citep{hojsgaard2008graphical}. Most of the methods mentioned in Section \ref{sec:methods} can be applied to the coloured GGMs. For example, \cite{colouredgraphs1} implemented an RJ method by utilizing the colored $G$-Wishart prior as the Diaconis-Ylvisaker conjugate prior for GGMs. \cite{roverato2022model} applied a stepwise model search procedure to estimate a brain network from fMRI data. 

\textbf{Time-varying GGMs}: 
While extensive literature revolves around learning static graphical time-invariant models, the change of interdependencies with a covariate (\textit{e.g.}, time or space) is often the rule rather than the exception for real-world data, such as friendships between individuals in a social community, communication between genes in a cell, equity trading between companies, and computer network traffic.
Bayesian approaches in time-varying GGMs are computationally burdensome with time complexity. Thus, 
\cite{yu2022efficient} developed a low-complexity Bayesian approach by imposing temporally dependent spike-and-slab priors on the graphs such that they are sparse and vary smoothly across time.

\textbf{Multivariate regressions}: 
GGMs, and sparse graphical models more broadly, are applicable for statistical inference in high-dimensional regression. This application is straightforward when the response variables and explanatory variables are collectively viewed as a set of $p$-dimensional variables within our graphical models, as discussed in Section 6 of \citet{Stochastic2013}, for example. In such instances, the inference process can address both the selection of relevant variables and the quantification of uncertainty for the regression coefficients, as outlined in the works of \citet{liang2022markov, liang2013bayesian}.



\subsection{Future Research}
In the coming decade, Bayesian methods have the potential to become even faster, expand their theoretical guarantees, grow beyond the Gaussian case, and make more impact with their applications. This section speculates how.

First, new MCMC methods can increase the efficiency and feasible dimension of Bayesian structure learning. \citet{MPLpaper} show the potential of MCMC methods that move only over the graph space, and not over the space of precision matrices. These methods, however, still only allow graphs to change at most one edge per MCMC iteration. It is an open question whether the balance conditions still hold when allowing changes of multiple edges. If true, this could lead to a significant reduction in computation time. Similarly, \cite{WWA} show the benefit of the informed proposal and delayed acceptance techniques for the reversible jump approach on the joint space of graphs and precision matrices. It remains to be shown whether methods on the graph space can benefit from the same techniques. \citet{SSWang} introduces the spike-and-slab prior in Bayesian structure learning. He circumvents the calculation of the normalizing constant by putting its inverse in the prior of the graph. This leads to a computationally efficient method but creates a challenge when incorporating any knowledge of the graph structure in the prior of the graph. This raises two questions: (i) Could the normalizing constant of the spike-and-slab prior be approximated? (ii) Could $G$-Wishart methods on the joint space benefit from a similar trick? 

Second, moving away from the standard MCMC approach offers a promising and unexplored perspective to Bayesian structure learning. \citet{SMCreview} argue that Sequential Monte Carlo (SMC) methods remain under-used in statistics, despite several advantages. \cite{SMCBoom} and \cite{SMCTan} show that the SMC approach works for Bayesian structure learning. However, their SMC methods cannot yet compete in terms of computational efficiency with the MCMC methods. Likewise, \citet{Gradient} outline the benefits of the stochastic gradient MCMC (SGMCMC) method, and \cite{GradientExample} design an SGMCMC method for exponential random graph models. As of now, however, no SGMCMC structure learning method exists. 

Third, Bayesian structure learning needs more theoretical results. Although the majority of MCMC methods meet the balance conditions, not one of them provides rates of convergence as in \citet[Section 3.2]{stationary_distr_exists}. Moreover, among the methods in Table \ref{tab:searchliterature}, only one \citep{SMCBoom} provides an unbiased estimate. In the future, more theoretical results are essential to the credibility of Bayesian structure learning.

Fourth, recent improvements in structure learning in GGMs could enhance methods beyond the general Gaussian case. The most apparent is the non-Gaussian case, in which the state-of-the-art methods discussed in this paper can be directly applied. This avenue, however, is barely investigated. Similarly, other related fields can benefit from the recent strides made in the general Gaussian case. They include multiple Gaussian graphical models \citep{SMCTan}, coloured Gaussian graphical models \citep{colouredgraphs1}, and graphical models with external network data \citep{networkdata}.

Lastly, the increase in the feasible dimension of Bayesian methods will enhance their applications. In Section \ref{sec:application}, we reduce the data set containing several thousands of genes to a mere one hundred genes to make the data feasible for computation, potentially losing valuable information. This kind of dimension reduction will be decreasingly necessary as Bayesian methods improve. Likewise, in neuroscience, the models of the brain no longer have to be simplified to just $100$ areas \citep{Alzheimer}. This could potentially improve the understanding of cognitive diseases like Alzheimer's. The enhancements in Bayesian structure learning also open up new applications. Especially exciting but yet unexplored examples are graph neural networks and large language models, which make use of the dependency networks among a large number of variables.

\section*{Acknowledgements}
The authors thank the Editor, Prof. Lee, the Associate Editor, and the two referees for their insightful comments, which have significantly improved this article.

\section*{Disclosure}
The authors report there are no competing interests to declare.

\bigskip
\begin{center}
{\large\bf SUPPLEMENTARY MATERIAL}
\end{center}

\begin{description}

\item[GitHub repository:] All scripts used to perform the simulation study of Section \ref{sec:simulation} and the case study of Section \ref{sec:application} can be found on \sloppy \url{https://github.com/lucasvogels33/Review-paper-Bayesian-Structure-Learning-in-GGMs}.
\item[Supplementary materials:] The supplementary materials provide (S1) the references to the code of all algorithms compared in Section \ref{sec:simulation}, (S2) all algorithm-specific settings for the simulation study of Section \ref{sec:simulation}, (S3) the data dictionary for the data used in Section \ref{sec:application}, (S4) boxplots showing prior sensitivity of the BD-A and SS-O algorithms on the human gene data set in Section \ref{sec:application}, (S5) a table to convert the gene numbers of Figure \ref{fig:gene_network} to the names of the genetic transcripts.
\end{description}
 
\bibliographystyle{agsm}
\bibliography{sampleBIB}
\end{document}


\maketitle

\section{Source of algorithm codes}
In Section 4 ("Empirical Comparison") we compare 8 algorithms. Table \ref{tab:code_source} shows for each of these algorithms where we obtained the code and in which language it was originally written.  

\begin{table}[h]
\centering
\begin{tabular}{lllll}
\toprule
\textbf{\#} & \textbf{Algorithm} & \textbf{Source type} & \textbf{Reference} 
& \textbf{Code} 
\\
\hline 
1                  & glasso                   & R package \textit{Huge} & \citet{huge_package}                                                                & R, C++        \\
2                           & RJ - WWA                 & Github page & \citet{github_RJWWA} & R, C++        \\
3                           & BD - A                   &  R package \textit{BDgraph} & \citet{BDgraph}                                                 & R, C++        \\
4                           & SS - O                   & R package \textit{ssgraph} & \citet{ssgraph}                                                             & R, C++        \\
5                           & ECM - BAGUS               & .zip file & \citet[Supp. mat.]{SSLasso}                                        & R             \\
6                            & G - MPLBD                & R package \textit{BDgraph} & \citet{BDgraph}                                                              & R, C++        \\
7                            & K - Horseshoe            & Github page & \citet{github_horseshoe}                        & Matlab            \\
8                   & H - BGGM                 & R package \textit{BGGM} & \citet{BGGM}                                                                & R, C++    \\
\bottomrule
\end{tabular}
\label{tab:code_source}
\caption{\textit{Every algorithm of Section 4 ("Empirical comparison") with their corresponding source type, reference and coding language.}}
\end{table}

\noindent For most algorithms we had to slightly alter the code to make it fit in our simulation framework. The K - Horseshoe was originally written in Matlab. We rewrote the code to R. The final code that was used to produce the results in Section 4 is available on the Github page: \url{https://github.com/lucasvogels33/Review-paper-Bayesian-Structure-Learning-in-GGMs}

\section{Algorithm specific settings}
In Section 4 ("Empirical Comparison") we compare 8 algorithms. Table \ref{tab:method_specs} shows for each of these algorithms what settings where used. All computations were performed on the University of
Amsterdam’s supercomputer (Snellius system) using one core on a Lenovo® ThinkSystem
SR645 2.60 GHz Processor.

\newgeometry{bottom=20mm,top=20mm,left=15mm,right=15mm}

\begin{landscape}

\begin{threeparttable}[]
\centering
\rowcolors{2}{gray!40!lightgray!50}{gray!10!lightgray!100}

\begin{tabular}{lllll}
\toprule
\textbf{Algorithm} & \textbf{Tuning method}                                                                                                                      & \textbf{\begin{tabular}[c]{@{}l@{}}Prior density  \\ on the graph\end{tabular}} & \textbf{MCMC settings}                                                                                                                                                                                                    & \textbf{Other settings}   
\\ \hline
 
glasso                     & \begin{tabular}[c]{@{}l@{}}Rotation information \\ criterion  \tnote{a}\end{tabular}  

& NA                                                                     & NA                                                                                                                                                                                                                        & \begin{tabular}[c]{@{}l@{}}Algorithm stops \\ when converged \tnote{b} \end{tabular}           \\
 
RJ - WWA                   & NA                                                                                                                                   & 0.2                                                                    & \begin{tabular}[c]{@{}l@{}}burnin = 0\\ MCMC iterations p =10: 30000 \\ MCMC iterations  p=50: 50000 \\ initial graph:   empty graph\end{tabular}                                                                         & \begin{tabular}[c]{@{}l@{}}With delayed acceptance\\ Without informed proposal \\ With approx. of the norm. constants \\ of \citep{Accelerating} \end{tabular} \\
 
BD - A                     & NA                                                                                                                                   & 0.2                                                                    & \begin{tabular}[c]{@{}l@{}}burnin = 0\\ MCMC iterations p =10: 30000 \\ MCMC iterations  p=100: 50000\\ Initial graph:  empty graph\\ Initial precision matrix: empty matrix\end{tabular}                                 &                                                                                                                                                               \\
 
SS - O                     & \begin{tabular}[c]{@{}l@{}}No tuning, instead we set \\ $v_0 = 0.02$,  $v_1 = 2$ and $\eta = 2$\end{tabular}                                                                                                                                      & 0.2                                                                    & \begin{tabular}[c]{@{}l@{}}burnin = 0\\ MCMC iterations p =10: 3000 \\ MCMC iterations p =100: 3000 \\ MCMC iterations p =1000: 100 \\ Initial graph: empty graph\\ Initial precision matrix: empty matrix\end{tabular}   &                                                                             \\
 
ECM - BAGUS                 & \begin{tabular}[c]{@{}l@{}}Bayesian information\\ criterion, as in \\ \citep[Section 4.4]{SSLasso}\end{tabular}                                                                                                            & 0.5                                                                    & NA                                                                                                                                                                                                                        & Initial precision matrix: identity matrix                                                                                                                     \\
 
G - MPLBD                  & NA                                                                                                                                   & 0.2                                                                    & \begin{tabular}[c]{@{}l@{}}burnin = 0\\ MCMC iterations p =10: 30000\\ MCMC iterations p =100: 30000\\ MCMC iterations p =1000: 200000\\ Initial graph: empty graph\\ Initial precision matrix: empty matrix\end{tabular} &                                                                                                                                                               \\
 
K - Horseshoe              & NA                                                                                                                                   & NA                                                                     & \begin{tabular}[c]{@{}l@{}}burnin = 0\\ MCMC iterations p = 10: 2000 \\ MCMC iterations p =100: 2000 \\ MCMC iterations p =1000: 100 \\ Initial precision matrix: identity matrix\end{tabular}                            & \begin{tabular}[c]{@{}l@{}}Point estimate of the graph calculated \\ using 50\% credible intervals. \\
Initial parameter settings \\ as in \citep{GraphHorseshoe} \end{tabular}                                                            \\
 
H - BGGM                   & NA                                                                                                                                   & NA                                                                    & NA                                                                                                                                                                                                                        & \begin{tabular}[c]{@{}l@{}}All settings are the default \\ of the BGGM R package \\ \citep{BGGM} \end{tabular}                                                              \\   \bottomrule
\end{tabular}

\begin{tablenotes}
\item[a]{\small \textit{Using the rotation information criterion, the best $\lambda$ is selected from a vector of length $30$ that ranges evenly on a log scale from $\lambda_{max}$ to $\lambda_{min}$, where $\lambda_{max}$  is the lowest value for which all entries in $K$ are zero and $\lambda_{min} = 0.01\lambda_{max}$.}}
\item[b]{\small \textit{Algorithm is considered converged when the average absolute parameter change is less than 0.0001 times the average absolute value of the off diagonal elements of $K$.}}
\end{tablenotes}

\label{tab:method_specs}
\caption{\textit{Every algorithm of Section 4 ("Empirical comparison") with their corresponding tuning method, the prior density of the graph, the MCMC settings and other settings.}}

\end{threeparttable}

\end{landscape}

\restoregeometry

\section{Data dictionary}

For Section 5 ("Case Study") we used a gene dataset. The data are collected from $n = 60$ unrelated individuals. For each individual the $p = 100$ most variable genetic transcripts were selected. The resulting data is a $60 \times 100$ matrix $\mathbf{X}$, whose elements $x_{ij}$ denote the level of protein corresponding to gene $i = 1,2..,p$ measured in individual $j = 1,2,..,n$. The column names of the matrix are unique identifiers of the $60$ individuals. They start with \textit{NA} and end with $5$ digits. For example \textit{NA06985}. The row names of the matrix are unique identifiers of the $100$ genetic transcripts and are listed in Section \ref{sec:gene_table} of this supplementary material. For more information on the collection, cleaning and preparation of these data, please see \citet[Section 4]{Bhadra2013}.

The data are accesible in the supplementary material of \citet{Bhadra2013} or by using the function geneExpression() in the R package BDgraph \citep{BDgraph}.

\section{Prior sensitivity}
For the data application in Section 5, we selected a prior density of $20\%$. The impact of this prior density on the posterior remains relatively unexplored in Bayesian structure learning literature. In order to test the sensitivity of the algorithms to the priors, we ran the algorithms with a sparse ($1\%$ density), average ($20\%$ density) and dense ($50\%$ density) graph prior. We compare the boxplots of the resulting edge inclusion probabilities in Figure S1 (SS-O algorithm) and in Figure S2 (BD-A algorithm). As expected, a higher prior density results in higher edge inclusion probabilities with the SS-O algorithm being slightly more sensitive to a higher prior density. For all prior densities, however, both algorithms are able to distinguish between edges with a low and high edge inclusion probability.

\begin{figure}[H]
     \centering
     \includegraphics[width=0.99\linewidth]{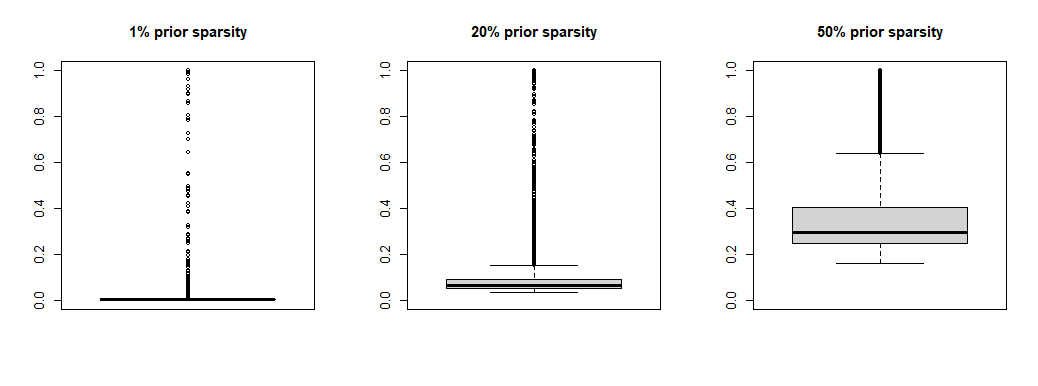}
    \caption{\textit{Boxplots of the final edge inclusion probabilities of the SS-O algorithm.}}
    \label{fig:ss_prior_sens}
\end{figure}

\begin{figure}[H]
     \centering
     \includegraphics[width=0.99\linewidth]{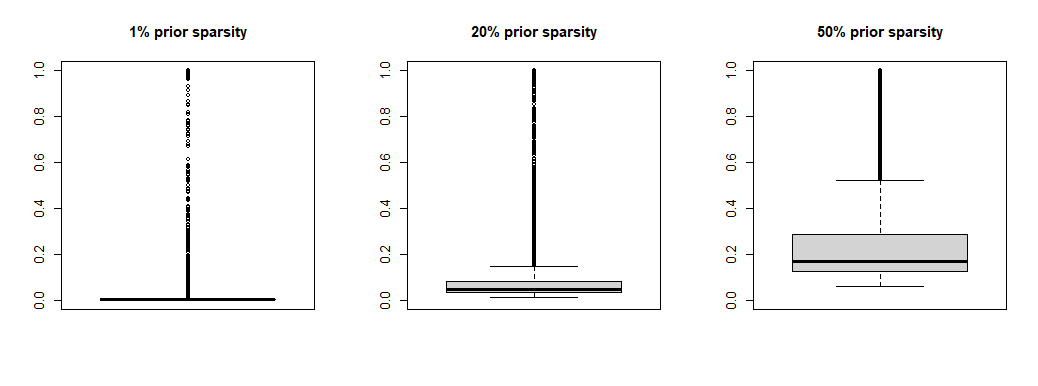}
    \caption{\textit{Boxplots of the final edge inclusion probabilities of the BD-A algorithm.}}
    \label{fig:bd_prior_sens}
\end{figure}

\section{Gene conversion table}
\label{sec:gene_table}

\begin{table}[H]
\begin{tabular}{ |l  l |l| l l|}
\arrayrulecolor{blue}
\cline{1-2} \cline{4-5}
\textbf{Node nr.} & \textbf{Gene name}& & \textbf{Node nr.} & \textbf{Gene name}\\
\cline{1-2} \cline{4-5}
1 & GI\_18426974-S && 51&GI\_24308084-S \\
 2 & GI\_41197088-S && 52&GI\_19743804-A \\
 3 & GI\_17981706-S && 53&GI\_28610153-S \\
 4&GI\_41190507-S && 54&GI\_27764881-S \\
 5&GI\_33356162-S && 55&GI\_4507820-S \\
 6&Hs.449605-S && 56&GI\_21464138-A \\
 7&GI\_37546026-S && 57&GI\_14211892-S \\
 8&Hs.512137-S && 58&GI\_27894329-S \\
 9&GI\_37546969-S && 59&Hs.185140-S \\
 10&Hs.449584-S && 60&GI\_4505888-A \\
 11&Hs.406489-S && 61&GI\_34222336-S \\
 12&GI\_18641371-S && 62&GI\_27477086-S \\
 13&Hs.449609-S && 63&GI\_4504436-S \\
 14&hmm3574-S && 64&GI\_21614524-S \\
 15&hmm10298-S && 65&GI\_24497529-I \\
 16&Hs.449602-S && 66&GI\_19224662-S \\
 17&GI\_11095446-S && 67&GI\_22027487-S \\
 18&Hs.512124-S && 68&GI\_9961355-S \\
 19&GI\_37542230-S && 69&GI\_21389558-S \\
 20&hmm3577-S && 70&GI\_24797066-S \\
 21&GI\_21389378-S && 71&GI\_38569448-S \\
 22&GI\_27754767-I && 72&GI\_28557780-S \\
 23&GI\_13514808-S && 73&GI\_20302136-S \\
 24&GI\_13027804-S && 74&GI\_16159362-S \\
 25&GI\_4504410-S && 75&Hs.171273-S \\
 26&GI\_11993934-S && 76&Hs.136376-S \\
 27&GI\_33356559-S && 77&GI\_4502630-S \\
 28&GI\_37537697-S && 78&GI\_4504700-S \\
 29&hmm10289-S && 79&GI\_7657043-S \\
 30&GI\_42660576-S && 80&GI\_42476191-S \\
 31&GI\_34915989-S && 81&GI\_37546229-S \\
 32&GI\_31377723-S && 82&GI\_7662241-S \\
 33&GI\_28416938-S && 83&GI\_13325059-S \\
 34&GI\_42656398-S && 84&GI\_4505948-S \\
 35&Hs.449572-S && 85&GI\_41350202-S \\
 36&GI\_23065546-A && 86&GI\_20373176-S \\
 37&GI\_42662536-S && 87&hmm9615-S \\
 38&GI\_41190543-S && 88&GI\_21389548-S \\
 39&GI\_23510363-A && 89&hmm3587-S \\
 40&GI\_7661757-S && 90&GI\_7019408-S \\
 41&GI\_27482629-S && 91&GI\_37537705-I \\
 42&GI\_16554578-S && 92&GI\_31341400-S \\
 43&GI\_34222299-S && 93&GI\_18641372-S \\
 44&GI\_31652245-I && 94&GI\_30409981-S \\
 45&GI\_27754767-A && 95&GI\_5454143-S \\
 46&GI\_8923472-S && 96&GI\_40354211-S \\
 47&GI\_20070269-S && 97&GI\_18379361-A \\
 48&GI\_30795192-A && 98&GI\_20143949-A \\
 49&GI\_31077201-S && 99&GI\_4507426-S \\
 50&GI\_27894333-A && 100&GI\_14602456-I \\
\cline{1-2} \cline{4-5}
\end{tabular}
\caption{\textit{Names of genetic transcripts corresponding to node numbers of Figure 5. For more information on the 100 genetic transcripts, see } \citet{Bhadra2013}}

\end{table}

\bibliographystyle{agsm}
\bibliography{suppmatbib}